\documentclass[aip,preprint]{revtex4-1}

\usepackage[separate-uncertainty=true,number-mode=math,unit-mode=text]{siunitx}
\usepackage{graphicx}
\usepackage{amsfonts,amsmath,amssymb,upgreek}
\usepackage{bm}
\usepackage{wasysym} 
\usepackage{physics}
\usepackage{dcolumn}
\usepackage{xcolor}

\usepackage[utf8]{inputenc}
\usepackage[T1]{fontenc}
\usepackage{mathptmx}
\usepackage{etoolbox}

\usepackage{hyperref}
\hypersetup{
	colorlinks=true, 
	linkcolor={red!65!black},
	citecolor={blue!75!black},
	urlcolor={blue!50!black}
}

\makeatletter
\providecommand\add@text{}
\newcommand\tagaddtext[1]{%
	\gdef\add@text{#1\gdef\add@text{}}}%
\renewcommand\tagform@[1]{%
	\maketag@@@{\llap{\add@text\quad}(\ignorespaces#1\unskip\@@italiccorr)}%
}
\makeatother

\draft 

\begin{document}
	
	
	\title{Merging Two Molecular Beams of ND$_3$ up to the Liouville Limit} 
	
	
	
	\author{S.\,E.\,J. Kuijpers}
	\author{A.\,J.\,A. van Roij}
	\author{E. Sweers}
	\author{S. Herbers}
	\author{Y.\,M. Caris}
	\author{S.\,Y.\,T. van de Meerakker}
	\email[Author to whom correspondence should be addressed: ]{b.vandemeerakker@science.ru.nl}
	\affiliation{Radboud University Nijmegen, Institute for Molecules and Materials, Heijendaalseweg 135,
		6525 AJ Nijmegen, The Netherlands}
	
	
	\date{\today}
	
	\begin{abstract} 
		In low-energy collisions between two dipolar molecules, the long-range dipole-dipole interaction plays an important role in the scattering dynamics. Merged beam configurations offer the lowest collision energies achievable, but they generally can not be applied to most dipole-dipole systems as the electrodes used to merge one beam would deflect the other. This paper covers the design and implementation of a merged electrostatic guide whose geometry was numerically optimized for ND$_3$-ND$_3$ and ND$_3$-NH$_3$ collisions. This device guides both beams simultaneously and makes them converge up to an effective collision angle of \SI{2}{\degree}, yielding the optimal compromise between spatial overlap and the lowest possible collision energy. We present preliminary data for inelastic ND$_3$-ND$_3$ collisions.
	\end{abstract}
	
	
	\maketitle 
	
	\section{Introduction}
	
	The study of molecular collisions in crossed molecular beam experiments at energies well below 1 kelvin has emerged as a new and exciting research frontier. Yet, designing experiments to reach sufficiently low collision energies (and sufficienty high energy resolutions) has proven a challenging task. In recent years, several novel experimental approaches to study low-energy collisions have been developed and successfully implemented, and a variety of new methods are being pursued that hold great promise to accomplish this feat in the future. Low interaction energies have been achieved by using a small beam crossing angle~\cite{Chefdeville:Science341:06092013,Amarasinghe2020} or by controlling the reagent's velocity using a decelerator~\cite{Meerakker:CR112:4828}. The lowest collision energies are reached when the relative velocity between both collision partners approaches zero. This can be achieved using intrabeam collisions~\cite{Amarasinghe2017,Perreault:science358:356,Gawlas2019,Dulitz2020,Smith2006}, i.e., by inducing collisions between atoms and molecules that are copropagating in a beam originating from a single source, or by merging two individual molecular beams by tangentially overlapping one beam on the other beam's path~\cite{Henson:Science338:234,Gordon2018,Zhelyazkova2020,Tang2023}. Typically, merging is achieved using beam manipulation tools that use the force induced by inhomogenuous magnetic or electric fields employing the Zeeman or Stark effect, respectively.     
	
	Beam merging is a well-known technique in collision studies involving ion beams~\cite{Phaneuf1999,OConnor2015}, but its successful application in collision experiments with neutrals has only been reported recently. The merged beam approach is particularly appealing as the two beams containing the collision partners are produced independently, allowing for collision studies between a large number of atomic and molecular species within a large range of collision energies. However, this technique hinges on a key property of the particles in the two beams: they must have a different interaction with the force that merges the beams. Ideally, particles in one of the beams should be  completely inert to the merging force, such that particles in the other beam can be manipulated and overlapped without affecting the first beam's path. 
	
	In successful merged beam implementations thus far, this problem was circumvented using an appropriate choice of system. Narevicius and coworkers merged beams of metastable He or Ne atoms with beams of Ar or H$_2$ using a magnetic quadrupole guide~\cite{Henson:Science338:234,Lavert-Ofir:NatChem6:332,Klein:NatPhys13:35,Margulis2022}. Whereas the unpaired spins of the metastable atoms offered a significant magnetic dipole moment for efficient manipulation, the singlet electronic states of Ar and H$_2$ guaranteed complete insensitivity to magnetic fields. Osterwalder and coworkers developed a dual-guide merged beam experiment, in which beams of molecules with an electric dipole moment were overlapped with beams of atoms with a magnetic dipole moment using curved electric and magnetic focusers, respectively~\cite{Osterwalder:EPJ-TI2:10}. Recently, Tang \emph{et al.} reported on a merged beam experiment using beams of NO and ND$_3$ that were merged by bending the ND$_3$ molecules into the NO beam's path using an electrostatic curved hexapole~\cite{Tang2023}. The much smaller dipole moment of NO compared to ND$_3$ ensured that the deflection of NO by the curved hexapole was negligible. Beam merging using miniaturized electrode designs has also been achieved using Rydberg states that offer extremely large electric dipole moments~\cite{Hogan2016,Zhelyazkova2020}.         
	
	The implementations mentioned above unfortunately exclude an important and interesting class of systems: the scattering of two molecules that each possess a significant (electric) dipole moment. These systems are of particular interest, as the long-range and anisotropic dipole-dipole interaction offers an exquisite `handle' to control and steer collisions~\cite{Koller2022}. At sufficiently low temperatures, small external electromagnetic fields can vastly change the interaction, yielding distinctive opportunities to engineer interaction Hamiltonians and to steer collision pathways. In addition, scattering of two identical molecules offers a wealth of opportunities to study resonant energy transfer using (near)-matching energy levels. Recently, a peculiar and highly interesting scattering mechanism was predicted to occur in low-energy collisions between dipolar molecules that possess near-degenerate rotational energy levels with opposite parity\cite{Tang2023}. A local maximum in the cross section was predicted at a collision energy that connects to the Langevin region at high energies and the Wigner regime at low energies. Collisions involving molecules such as OH, ND$_3$, H$_2$CO, NH ($a\,^1\Delta$), CH, SH and CO ($a\,^3\Pi$) all possess a significant dipole moment and the required energy level structure, and are therefore prime candidates for low-energy scattering experiments.  
	
	Merging two dipolar molecules with near-matching velocities appears fundamentally impossible, however, as the beam manipulation tools required to steer one beam will necessarily also affect the other. This problem is more formally expressed as Liouville's theorem, which states that the phase space density of a Hamiltonian system remains constant along every trajectory~\cite{Liouville:JMPA3:342,Ketterle:PRA46:4051}. For two beams of identical particles, the `system' encapsulates both beams. Since the phase space density has to remain constant, one cannot overlap (i.e. merge) the two separate areas of the phase space, as this would increase the density. Merging (overlap in the spatial coordinate) is then only possible by allowing the packets to heat up (non-overlap in the velocity coordinate), setting a lower limit for the collision energy that can be obtained. This situation is also found in trapping experiments: simply adding a new packet of molecules to an existing sample of trapped molecules is not possible without heating or losing the already trapped molecules.   
	
	Certainly, for specific systems, one could contemplate strategies to circumvent this `Liouville limit' making smart use of the energy levels of a molecule as its dipole moment can effectively vanish for certain electronic or rotational energy levels. For instance, NH ($a\,^1\Delta$) radicals or CO ($a\,^3\Pi$) molecules are easily manipulated using electric fields due to the pronounced Stark effect by virtue of close lying $\Lambda$-doublets, whereas NH ($X \, ^3\Sigma^-$) and CO ($X ^1\Sigma^+$) are immune to electric fields~\cite{Riedel:EPJD65:161,Engelhart2015}. Laser-induced population transfer from one electronic state to the other can therefore in principle be used to effectively switch off the dipole moment just before merging, and to switch it on again thereafter. Similar ideas may be considered exploiting the Stark splitting of rotational energy levels in an external electric field according to their quantum number $M$. Molecules that possess states with $M=0$ experience (at least in first order) no Stark effect, and population transfer to a state with $M=0$ may allow for perfect merging. For instance, ND$_3$ can be prepared in the $M=0$ component of the $J_K^p=1_1^+$ lower inversion-doublet component of the rotational ground state via a microwave induced $1_1^-, |M|=1 \rightarrow 1_1^+, M=0$ transition in the presence of an electric field~\cite{Herbers2022} (throughout this paper we follow the common convention that $J$ refers to the rotational angular momentum of the molecule, $K$ refers to the projection of $J$ on the symetry axis of the molecule, and $M$ refers to the projection on the laboratory frame Z-axis (parallel to the applied electric field). The label $p$, either $+$ or $-$, indicates the parity of the tunneling split wavefunction at no electric fields applied.). Alternatively, the $2_1^+, M=0 \leftarrow 1_1^-, |M|=1$ transition can be induced using \si{THz} radiation at a frequency around \SI{600}{GHz}~\cite{Herbers2024}. Finally, some molecules that have both an electric and magnetic dipole moment have the interesting property to lose their magnetic dipole moment when prepared in a given electronic state. The OH radical is an interesting example within this category. In its $X\,^2\Pi_{3/2}$ electronic ground state, OH has a significant electric and magnetic dipole moment~\cite{Sawyer:PRL98:253002}, where in the $X\,^2\Pi_{1/2}$ spin-orbit excited state only the electric dipole moment remains. This may allow for the merging of OH ($X\,^2\Pi_{1/2}$) and OH ($X\,^2\Pi_{3/2}$) radicals using a magnetic guide.
	
	Yet, despite the fundamental restrictions imposed by the Liouville theorem and potential work-arounds, one may still consider experimental designs that spatially overlap particles originating from two separate molecular beams, and accept the consequence that the lowest collision energy achievable is compromised. This can be accomplished by merging two electrostatic guides carefully, as has been demonstrated for miniaturized structures on a chip~\cite{Deng2007,Deng2011,Palmer2017}. Alternatively, novel structures can be designed where a single electrode is shared by two independent beam manipulation elements, as demonstrated in a honeycomb hexapole structure by Shimizu \emph{et al.}~\cite{Shimizu2003}. Using a pioneering idea which was first introduced by Osterwalder and coworkers~\cite{Gordon2017:3D}, an electrostatic hexapole (6 electrodes) can gradually be converted into two quadrupole focusers of 4 electrodes each using two Y-shaped electrodes. The resulting `beamsplitter' was successfully used to split a beam of ND$_3$ molecules into two components. Later, the same group demonstrated the merging of two ND$_3$ beams by operating the beamsplitter in reverse~\cite{Tanteri:thesis:2021}. To date, however, no collision experiments have been reported using this device.       
	
	Here, we present a novel implementation of a merged guide, inspired by the guide presented by Osterwalder \emph{et al.}, that is specifically optimized for ND$_3$-ND$_3$ and ND$_3$-NH$_3$ collisions. We define quantitative criteria for the electrode design and implemented numerical optimization procedures to come as close as possible to the `Liouville limit'. We describe technical aspects of our design that ensures optimal high voltage compatibility. We experimentally tested the performance of the device using conventional and Stark-decelerated beams of ND$_3$ and NH$_3$, which compares favorably with the performance predicted by numerical trajectory simulations. We present preliminary collision data including near-matching velocities of both collision partners, demonstrating the feasibility of using this device for low-energy ammonia-ammonia collisions.
	
	This article is organized as follows: In Sec.~\ref{sec:Andreas} we summarize the original design by Osterwalder and coworkers. In Sec.~\ref{sec:design} we describe the criteria and underlying design concepts for our merged guide, and describe differences compared to the design used by Osterwalder \emph{et al.}. In Sec.~\ref{section:numerical_optimization} numerical optimization procedures are presented to tailor the design to reach optimal overlap in phase space for the two merging packets, i.e., strategies to achieve merging up to the Liouville limit. The mechanical implementation of our merged guide is described in Sec.~\ref{sec:implementation}, with particular emphasis on electrode production, mounting procedures, HV compatibility, and interfacing within our existing crossed beam setup involving a Stark decelerator. Experiments to test the performance of the merged guide using beams of ND$_3$ and NH$_3$ are presented in Sec.~\ref{sec:results}, and comparisons with predictions from numerical trajectory simulations are made. Finally, we present a preliminary collision experiment in Sec.~\ref{sec:collisions} that demonstrates the feasibility of low-energy merged beam collision experiments using our guide.   
	
	\section{Original idea and manufacturing}\label{sec:Andreas}
	
\begin{figure}
		\includegraphics[trim={0 0.2cm 0 0.6cm},clip]{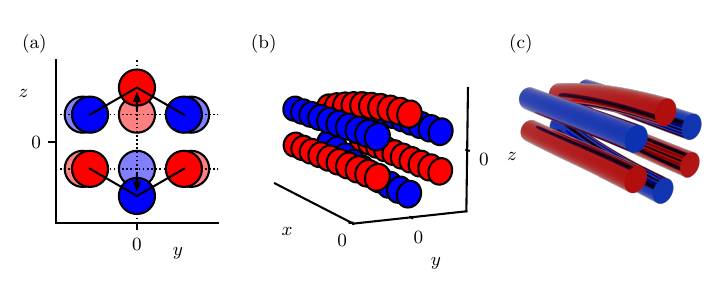}
		\caption{
			\label{fig:basic_merge}
			(a) Schematic illustration of how two adjacent quadrupoles can be merged into a single hexapole. (b) Sliced view of a merged guide, in which each ($y,z$) cross section is defined by its position along the $x$-axis. (c) 3D render of the merged guide.
		}
	\end{figure}

	Let us first describe the design by Osterwalder \emph{et al.}, as it served as inspiration for the merged guide we developed. Osterwalder and coworkers started with two independent quadrupoles that are gradually brought together~\cite{Gordon2017:3D}. If the quadrupole polarities are chosen such that they are mirrored, the two neighboring sets of electrodes from each quadrupole can ultimately merge using Y-shaped electrodes (see Fig.~\ref{fig:basic_merge}(a)). In this figure, the first half of the Y-shaped electrodes are omitted, since two ideal quadrupoles along arbitrary trajectories can be maintained during this phase. The beam merging effectively occurs when the two half-overlapped quadrupoles start to morph into a single hexapole, as shown in Fig.~\ref{fig:basic_merge}(a).
	
	The resulting $3\times2$ array of rods still form two ideal quadrupole wells, with a single set of electrodes shared by both quadrupoles. Next, the rods that just merged are moved outwards to gradually morph the geometry and potential into that of a single hexapole as illustrated in Fig.~\ref{fig:basic_merge}(a). In principle, this idea can be used to merge any n-pole with an m-pole to make an $(\text{n}+\text{m}-2)$-pole guide.
	
	To construct this intricate shape, Osterwalder and coworkers fabricated the merged guide using 3D printing techniques. It was found that direct 3D printing of metal leads to porous samples with a rough surface unsuitable for high voltage (HV) applications; an experience we gained ourselves as well. Instead, the guide was printed from plastic, which was electroplated afterwards to make the surface electrically conductive. This production method is extremely quick, versatile and unlimited in the complexity of the electrode shapes. However, the bulk plastic unfortunately has inferior HV and heat resistance properties compared to metal. Although successful application of an impressive voltage difference of \SI{22}{kV} between neighboring rods was reported, there is a severe risk of high voltage arcing or surface currents along the plastic support structure, damaging the setup. 
	
	Luckily, although the electrodes have a complex shape, it is still possible to manufacture them through conventional, subtractive methods like CNC milling. With this approach it is unfortunately not possible to produce the entire guide in one go. Instead, each electrode has to be manufactured separately in addition to a mounting system to electrically isolate, support and align the electrodes, adding complexity to the setup. Yet, this will be our approach, as optimal HV performance is essential to our experiments. In addition, we made substantial changes to the merged guide design to optimally suit low-energy collision experiments in combination with a Stark decelerator.
	
	\section{Design}\label{sec:design}
	In determining the precise shape of the electrodes, two main questions need to be addressed: (1) How can the geometry of the merged guide be defined, and how is it parametrized? (2) What parameters can be modified to optimize its performance? These questions will be addressed in the following sections. 
	
	\subsection{The original merged guide concept}
	Let us start by defining a right-handed Cartesian coordinate system in which the $z$-axis points vertically upwards while the $x$-axis points along the main beam propagation direction. The origin is placed at the center of the end of the merged guide, where the electrode geometry is that of a hexapole.
	
	To fully define the geometry of the merged guide, it is sufficient to define the shape of every consecutive ($y,z$) cross section or `slice' along the main propagation direction, as depicted in Fig.~\ref{fig:basic_merge}(b). This can be achieved most conveniently in the following way. First, the mirror symmetry between the top and bottom of the design should be recognized.\footnote{It is noted that the electrode shape is symmetric, but the applied voltages are anti-symmetric.} Next, one can impose that the distance between the geometric center of adjacent rods equals three times the rod radius, $r_\text{rod}$, ensuring that the electrode surfaces are always one rod radius apart.\footnote{The distance between the electrode surfaces will be slightly smaller when measured along the surface normal instead of this cross section.} With this restriction, we have to define the displacement of one rod only, as the other rods follow accordingly.
	
	Unless mentioned otherwise, we will use a rod radius $r_\text{rod}=\SI{2}{mm}$ throughout. In the following, we denote the rods positioned at $y=0$ that move in the $z$-direction `inner rods' (see Fig.~\ref{fig:basic_merge}(a)), whereas the rods that move in the $y$-direction are referred to as `outer rods'. The entire merged guide geometry is parameterized by a single function $z_\text{inner}(x)$ by writing for the position of the outer rods
	
	\begin{equation}
		\frac{y_\text{outer}}{r_\text{rod}} = \pm \sqrt{3^2 - \left(\frac{z_\text{inner}}{r_\text{rod}} -1.5\right)^2} \qquad \text{,} \quad \frac{z_\text{inner}}{r_\text{rod}} \in (1.5,3),
	\end{equation}
	
	\noindent with $y_\text{outer}$ the $y$-coordinate of the geometric center of the outer rods and $z_\text{inner}$ the $z$-coordinate of the inner rod.
	
	Optimizing $z_\text{inner}(x)$ is crucial to maximizing the transmittance throughout the merged guide and overlapping the two beams with minimal collision energy, i.e., with minimal slushing of both packets when they evolve from the ideal quadrupole trapping field into the hexapole geometry. As a starting point let us assume a quadratic function, $\frac{z_\text{inner}}{r_\text{rod}}(x)=a-bx^2$, ensuring that the merging starts relatively fast but ends smoothly. For a merging segment of length $L$, the curve should pass through the points $(-L,1.5)$ and $(0,3)$, yielding $a=3$, $b=\frac{1.5}{L^2}$. This curve was used to draw the merged guide shown in Fig.~\ref{fig:basic_merge}(b,c).
	
	\begin{figure}[b!]
		\includegraphics{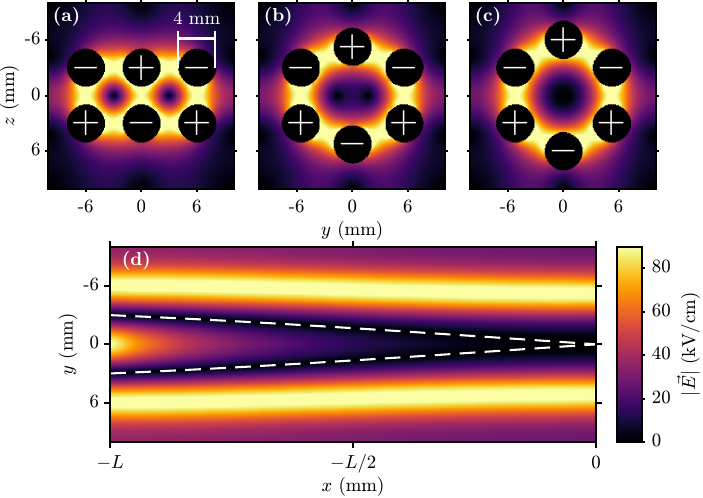}
		\caption{
			\label{fig:basic_merge_fields}
			Electric field strength along several cross sections of the merged guide, with \SI{\pm10}{kV} applied to its \SI{4}{mm} diameter rods. (a) ($y,z$) cross section at $x=-L$ showing the dual quadrupole well at the start of the merging process. (b) ($y,z$) cross section at $x=-L/2$ showing the middle of the merged guide. The barrier and distance between the two wells have decreased. (c) ($y,z$) cross section at $x=0$ showing the hexapole well once merging has been completed. (d) ($x,y$) cross section along the center of the merged guide, $z=0$, showing the guiding wells along the full length of the guide. The dashed white curves mark the guide minima. The color scale is identical for (a-d).
		}
	\end{figure}
	
	After drawing the merged guide in Autodesk Inventor, the 3D electrode geometry was imported into SIMION~\cite{Dahl2000} to calculate the position dependent electric field throughout the guide. The spatial resolution of SIMION was set to \SI{0.1}{mm}. The field strength along several cross sections of the merged guide is shown in Fig.~\ref{fig:basic_merge_fields}. Panels (a-c) show cross sections at three different positions along the $x$-axis to illustrate how the two quadrupole wells gradually deform and merge into one hexapole field. In panel (d), the field strength for the ($x,y$) cross section at $z=0$ is shown, illustrating how the trapping well evolves in the longitudinal direction. It is seen that the two quadrupole wells merge into a single hexapole well smoothly.

	\subsection{Different design concepts} \label{section:design_changes}
	
	Before optimizing the exact shape of the merged guide, there are two changes to the merged guide concept we would like to make.	 
	
	The first relates to the evolution of the wells along the longitudinal direction. Fig.~\ref{fig:basic_merge_fields} shows that both field minima follow a curved path. For our application in scattering experiments, it is actually preferred that one of the beams propagates through the guide in a straight line, for a number of reasons. First, in our experiments one beam of molecules is prepared by a Stark decelerator. It is beneficial to keep this exceptionally well defined beam aligned with our detector as a reference, even when the merged guide is turned off. This will aid calibration of the new merged guide and any subsequent collision experiments. In addition, it will ease troubleshooting by compartmentalizing the complexity of the total experiment. Second, a straight beam path ensures transmittance at a wide range of velocities, enabling access to a large range of collision energies. Third, with a straight beam path, molecules like NO that have a smaller dipole over mass ratio than ammonia, can still be used as a scattering partner.
	
\begin{figure}
		\includegraphics{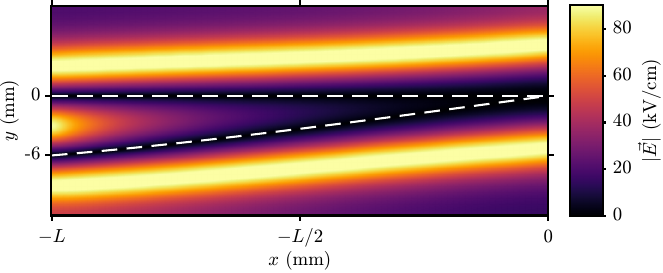}
		\caption{
			\label{fig:basic_merge_ywarp}
			Electric field strength like in Fig.~\ref{fig:basic_merge_fields}(d) after modifying the merged guide to straighten one of its paths. The white dashed curves mark the guide minima, showing that a straight trajectory is now possible.
		}
		
	\end{figure}

	So, to deflect the Stark decelerated beam as little as possible, one of the arms of the merged guide should follow the central axis of the Stark decelerator. This means that one well along every ($y,z$) cross section should be centered around the $z$-axis, which can be achieved by warping our previous design in the $y$-direction. From Fig.~\ref{fig:basic_merge_fields}(d) the positions of the field minima can be found directly. Fitting a third order polynomial bound by the position of the quadrupole well, $y_\textrm{Qwell} = 1.5r_\textrm{rod}$, on one end and the position of the hexapole well, $y_\textrm{Hwell} = 0$, on the other yields a smooth function that describes the required displacement for every ($y,z$) slice of the merged guide. Applying this warp to the previous design leads to a new merged guide, whose electric field is depicted in Fig.~\ref{fig:basic_merge_ywarp}. Note that one of the beams can now pass in a straight line, as desired.
	
	The second change we propose requires a bigger deviation from the original design. Up till now, to move the beams in each well closer together the center two rods had to shift outwards. This does not only bring the beams closer, but at the same time lowers the barrier between them. This can result in some particles already spilling over the barrier, instead of their trajectories being curved in the right direction. If it would be possible to maintain a high barrier while moving the minima closer together, the two beams would be separated better and they could be manipulated with a smaller radius of curvature. To achieve this the center electrodes can be made thinner.
	
\begin{figure}
		\includegraphics{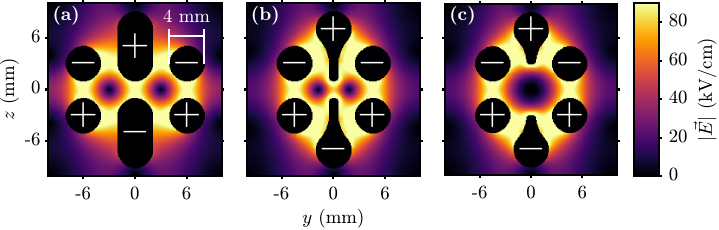}
		\caption{
			\label{fig:merge_finned_slice}
			Electric field strength along several cross sections of the finned merged guide, with \SI{\pm10}{kV} applied to the electrodes. (a) ($y,z$) cross section at the start of the merging, showing the dual quadrupole well. (b) ($y,z$) cross section at the middle of the merged guide, where the center electrodes are shaped like fins (c) ($y,z$) cross section once merging has been completed, showing the singular well. The color scale is identical for (a-c).
		}
	\end{figure}

	Decreasing the size of the center electrodes only works up to a certain width, as the electrodes will be prone to electrical discharge if their edges become too sharp. We assume a minimal radius of curvature of \SI{0.5}{mm} to be safe with respect to arcing, yielding a minimal width of \SI{1}{mm} for the center electrodes. The merging now takes place in two phases, as illustrated in Fig.~\ref{fig:merge_finned_slice}. First (panel a to b), the outer electrodes move inwards while the center electrodes decrease in width from \SI{4}{mm} diameter rods to \SI{1}{mm} wide `fins'. During this phase the distance between the two beams is reduced by almost a factor two, while the barrier between them stays intact. Second (panel b to c), the fins start moving outward, lowering the barrier and bringing the two beams in overlap. It should be noted that after the first phase, the outer rods are already closer to the center than in a standard hexapole. To correct for this and still make a confining potential, the fins do not retract completely. If the fins are retracted further than \SI{3}{mm} from the center plane we again form two potential wells, this time in the vertical $z$-direction. 
	
	With these design changes in mind, the next section will specify the exact dimensions of what we refer to as the finned merged guide. In Section~\ref{section:numerical_optimization}, the remaining free parameters will be optimized numerically to maximize the performance of the merged guide for our anticipated low-energy merged beam collision experiments.
	
	\subsection{The finned merged guide} \label{subsec:finned_merged_guide}
	Like before, the design of the finned merged guide can be fully specified by defining each ($y,z$) cross section along the $x$-direction. This time, however, we choose to parameterize the ($y,z$) cross section as a function of a new parameter $t$. The first and second phases of the merging process are described by the interval $t\in(-1,0)$ and $t\in(0,1)$, respectively. The geometry can later be optimized by tuning the position of each slice along the $x$-axis, which is captured by the function $x_\textrm{slice}(t)$.
	
	During the first phase, the outer rods move inwards linearly as a function of $t$ from $6$ to \SI{4.5}{mm}, and we can write
	
	\begin{equation}
		\label{eq:finned_yout}
		y_\text{outer}(t) = \pm (4.5 - 1.5\cdot t) \quad [\si{mm}] \qquad \text{,} \quad t \in (-1,0).
	\end{equation}
	
	\noindent At the same time, the width of the inner electrodes is reduced to a final value of \SI{1}{mm} in such a way that the distance between adjacent electrode surfaces stays again constant to one rod radius (\SI{2}{mm}). This ensures that the barrier which keeps the particles inside the guide stays intact, without creating areas of higher fields that might result in HV breakdowns. The distance between the resulting two vertical fins is also kept constant at \SI{2}{mm}.
	
	During the second phase the fins are retracted, whereas the positions of the outer electrodes remain fixed. The $z$-coordinate of the tip of the fins (i.e. the point closest to the ($x,z$) plane), $z_\textrm{tip}(t)$, increases from $z_\textrm{tip}(t=0)=\pm\SI{1}{mm}$ to $z_\textrm{tip}(t=1)=\pm\SI{3}{mm}$. To have the tip position converging towards the end of the merged guide, one additional constraint can be added on the slope of the curve
	
	\begin{equation}
		\left.\frac{dz_\textrm{tip}}{dt}\right\rvert_{t=1} = 0.
	\end{equation}
	
	\noindent These three constraints are satisfied by the second order polynomial
	
	\begin{equation}
		\label{eq:finned_ztip}
		z_\textrm{tip}(t) = \pm(1+4t-2t^2) \quad [\si{mm}] \qquad \text{,} \quad t \in (0,1).
	\end{equation}
	
	Eqs.~\ref{eq:finned_yout} and \ref{eq:finned_ztip} define the main features of the proposed merged guide geometry. To draw the complete design, one has to find similar equations that link points between the three panels of Fig.~\ref{fig:merge_finned_slice}, such that smooth 3D shapes can be drawn along the parametrized curves. Care should be taken to ensure that each electrode surface is smooth everywhere, especially in those regions where the fields are strongest. We will not discuss this process here, as these details hardly affect the fields along each beam, and hence the performance of the merged guide.
	
	Next, we again need to find an expression for the $y$-coordinate of the field minima, to ensure that one of the guiding arms aligns with the Stark decelerator axis. This will also later help us to construct a first guess for $x_\text{slice}(t)$ (see Sec.~\ref{section:numerical_optimization}). Once more, a polynomial with two boundary conditions can be used to fit the $y$-coordinate of the field minima in both phases of the merged guide. At $t=-1$ the two quadrupole wells are positioned at $y=\pm\SI{3}{mm}$. At $t=0$, the slice depicted in Fig.~\ref{fig:merge_finned_slice}(b), the field minima are positioned at $y=\pm\SI{1.77}{mm}$. Finally, at $t=1$ the merging is complete, leaving a single well at $y=0$. Taking this into account, fitting the field minima as a function of $t$ yields
	
	\begin{equation}
		\label{eq:y-warp}
		y_\text{well}(t) = \pm \begin{cases}
			1.77 - 1.26 \cdot t - 0.03 \cdot t^2 \quad [\si{mm}] \quad \text{,} \quad t \in (-1,0)   \\
			1.77 - 1.45 \cdot t - 0.32 \cdot t^2 \quad [\si{mm}] \quad \text{,} \quad t \in (0,1).  
		\end{cases}
	\end{equation}
	
	Now, after warping the design in the $y$-direction by Eq.~\ref{eq:y-warp}, one of the guiding arms again runs along the $x$-axis, while the other arm doubles in distance with respect to the $x$-axis. 
	
	Together, Eqs.~\ref{eq:finned_yout}, \ref{eq:finned_ztip} and \ref{eq:y-warp} define a family of merged guide designs characterized by their finned center electrodes, with the only remaining freedom to choose $x_\text{slice}(t)$. To ensure that the merged guide performs optimally in a certain experiment, $x_\text{slice}(t)$ can be numerically optimized specifically for the parameters pertaining to the experiment. This will be described in detail in the next section, focusing on the merging of two packets of ND$_3$ molecules. 
	
	\section{Numerical optimization} \label{section:numerical_optimization}
	We specifically optimize the finned merged guide for a planned experiment on low-energy ND$_3$-ND$_3$ collisions. This experiment can be regarded as the successor of a recent merged beam experiment using the NO-ND$_3$ system reported by Tang \emph{et al.} In that experiment, a packet of NO ($X \, {}^2\Pi_{1/2}, J=1/2f$) radicals was prepared using a Stark decelerator, and merged with a packet of ND$_3$ ($X \, ^1 \! A'_1, J_K^p=1_1^-$) molecules that was prepared using a curved hexapole. The curved hexapole had a bending radius of \SI{0.641}{m}, a bending angle of \SI{40}{\degree}, and ended a distance of \SI{65}{mm} before the laser detection region~\cite{Tang2023}. Packets of ND$_3$ with a fixed velocity of \SI{370}{m\per s} were guided by this curved hexapole by applying voltages of \SI{\pm12}{kV} to the hexapole rods. The collision energy was scanned by producing different velocities of the NO packet using the Stark decelerator. The relatively small dipole moment of \SI{0.16}{D} for NO compared to the \SI{1.5}{D} moment for ND$_3$ ensured that both beams could be merged without significantly affecting the NO packet for all velocities used. 
	
	To study low-energy ND$_3$-ND$_3$ collisions, the finned merged guide will replace the curved hexapole. We impose to the design that the distance between the end of the finned merged guide and the laser detection region remains \SI{65}{mm}, as this is the distance needed to incorporate a high-resolution velocity map imaging (VMI) detector~\cite{Plomp2020}. Traversing this distance, both packets should optimally be fully merged, i.e., we impose to the design that both beams in this region interact with the lowest possible collision energy and best spatial overlap. In addition, we also impose to the design that there is a minimal contribution of collisions that occur within the merged guide where both packets only partially overlap. This will ensure that we study collisions under well defined conditions, as these in-guide collisions occur in the presence of large inhomogenuous electric fields substantially complicating the interpretation of the scattering signals~\cite{Tang2023}. We furthermore assume a fixed velocity of \SI{370}{m\per s} for the curved arm of the merged guide, and that we can apply a maximum of \SI{\pm12}{kV} to the electrodes. 
	
	To assess the fitness of any $x_\text{slice}(t)$, the trajectories of 500 particles are simulated through the corresponding geometry. These particles have a mean velocity of \SI{370}{m\per s} and are initialized at the entrance of the curved arm of the merged guide with incident angle $\alpha$. Both the speed and direction of the particles are normally distributed, with a standard deviation of \SI{10}{m\per s} and \SI{0.5}{\degree}, respectively. Spatially, they are uniformly distributed along \SI{1}{mm}. In order to reduce the computational cost of these trajectory simulations the $z$-dimension is ignored, i.e. only particles that move in the ($x,y$) plane are considered.
	
	In order to numerically optimize $x_\text{slice}(t)$, two things are required: (1) A parametrization of $x_\text{slice}(t)$ with a reasonable initial guess. (2) An objective function, which evaluates the chosen $x_\text{slice}(t)$ with a single number or fitness. Optimization routines can then be used to vary the initial guess for $x_\text{slice}(t)$, and to find local minima of the objective function. This procedure is discussed below.
	
	\subsection{Parametrization and initial guess}
	Similar to Eq.~\ref{eq:y-warp}, we choose to parameterize $x_\text{slice}(t)$ as a polynomial for both phases of the merging. It should be a monotonically increasing function that is continuous everywhere and smooth except at $t=0$. We therefore write
	
	\begin{equation}
		\label{eq:x_polynomial}
		x_\text{slice}(t) = \begin{cases}
			a_1\cdot t^3 + a_2\cdot t^2 + a_3\cdot t + b_1 & \quad [\si{mm}] \quad \text{,} \quad t \in (-1,0) \\
			a_4\cdot t^3 + a_5\cdot t^2 + a_6\cdot t + b_2 & \quad [\si{mm}] \quad \text{,} \quad t \in (0,1).
		\end{cases}
	\end{equation}
	
	It should be noted that not all of these parameters are independent. To make this function continuous at $t=0$, $b_1=b_2$ should hold. Furthermore, to place the end of the merged guide at the origin, $x(1)=0$, one finds $b_2=-(a_4+a_5+a_6)$. This leaves 6 independent parameters, that can be represented as single vector, $\vec{a}=(a_1 , a_2 , a_3 , a_4 , a_5 , a_6)$. The starting angle of the particles, $\alpha$, is nonzero. It is either fixed at a value close to \SI{6}{\degree} or added to $\vec{a}$ as a seventh parameter.
	
	We start the numerical optimization with a reasonable initial guess, $\vec{a}_0$. This will make the algorithm converge faster, with less chance to end up in an irrelevant local minimum whose score is either too high, or whose corresponding solution is non-physical. To start with reasonable parameters, we pick $\vec{a}_0$ such that the curved arm follows a circle with radius $R=\SI{0.641}{m}$. From the equation of a circle with radius $R$, centered at $(0,-R)$, we obtain:
	
	\begin{align}
		R^2 &= x^2 + (R+y)^2 \\
		x &= -\sqrt{-2Ry-y^2}. \label{eq:x(t)_circle}
	\end{align}	
	
	\noindent Substituting $y=2y_\textrm{well}(t)$, with $y_\textrm{well}(t)$ given by Eq.~\ref{eq:y-warp}, yields an expression for $x(t)$. Fitting both phases of the finned merged guide with a third order polynomial yields our initial guess,
	
	\begin{equation}
		\vec{a}_0 = (0.84 \ , \ 4.47 \ , \ 23.79 \ , \ 57.28 \ , \ -37.83 \ , \ 38.41).
	\end{equation}
	
	\noindent Note that this fit is poor for $t\approx1$, where the slope of Eq.~\ref{eq:x(t)_circle} becomes infinite. This is not a problem since this is only a starting point for the optimization. 
	
\begin{figure}
		\includegraphics{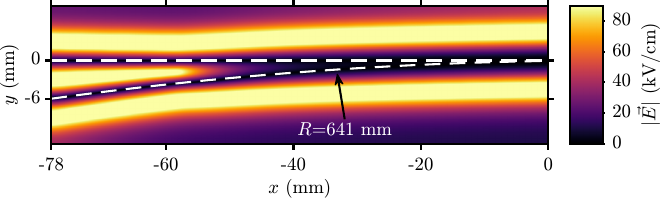}
		\caption{
			\label{fig:merge_circle}
			Electric field distribution within a $(x,y)$ cross section along the finned merged guide parameterized by initial guess $\vec{a}_0$. The white dashed lines show that the field minima follow a straight line and a perfect circle with a radius of \SI{641}{mm}.
		}
	\end{figure}

	Using $\vec{a}_0$ as initial parameters, the resulting electric field strength along the ($x,y$) cross section of the finned merged guide is shown in Fig.~\ref{fig:merge_circle}. As intended, one of the arms is aligned with the $x$-axis, while the other follows a circle with $R=\SI{0.641}{m}$. 
	
\begin{figure}
		\includegraphics{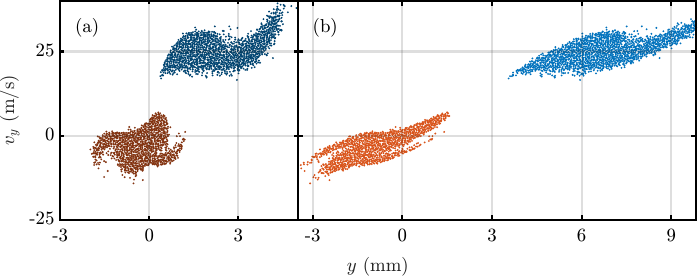}
		\caption{
			\label{fig:merge_x0_phasespace}
			$(y,v_y)$ phase space produced by the finned merged guide parameterized by initial guess $\vec{a}_0$. Particles guided through the curved arm are displayed as blue dots, while the particles guided through the straight arm are displayed as orange dots. (a) phase space at the exit of the merged guide. (b) phase space at the plane of detection, \SI{65}{mm} behind the exit of the merged guide.
		}
	\end{figure}

	To evaluate the performance of the design, it is instructive to look at the transverse ($y,v_y$) phase space distribution of both packets at various longitudinal positions along the beamline. These are displayed in Fig.~\ref{fig:merge_x0_phasespace}(a) as simulated from $\vec{a}_0$. Particles guided through the curved arm are displayed as blue dots, while the particles guided through the straight arm are displayed as orange dots. The dark colors represent the particles when they exit the merged guide, while the light colors in panel (b) represent the same particles at the plane of detection, \SI{65}{mm} behind the merged guide. As can be seen, this geometry actually performs poorly, as the particles guided through the curved arm have a large transverse velocity at the exit of the merged guide, even though they are roughly centered around $y=0$. This causes them to drift to positive $y$ values during the \SI{65}{mm} free flight further downstream, missing the point of detection and limiting the overlap with the collision partner. Clearly, the particles `fly off track' upon exiting the merged guide, and the curvature of the guide should be modified to prevent this. 
	
	\subsection{Objective function}
	Taking the poor performance of the initial guess as an example, it is clear that designing a structure that gradually converts two wells into a single well requires careful consideration. For our intended application, a proper merged guide should result in a transverse phase space distribution from the curved arm that is located as close as possible to the phase space from the straight arm, which is centered around $(y,v_y)=(0,0)$ (Note that perfectly overlapping both phase space distributions is forbidden by the Liouville theorem unless their combined phase space volume is maintained). To achieve a design through numerical optimization that meets this requirement, a single score for the performance of a geometry is required. An algorithm can in turn minimize this score by varying $\vec{a}$. The function that calculates this score is called the objective function, $f_\textrm{obj}$. Our requirements for $f_\textrm{obj}$ can be summerized as:
	
	\begin{enumerate}
		\item $f_\textrm{obj}$ is small $\iff$ many particles reach the detection laser, which we modeled with a diameter of \SI{1}{mm}.
		\item $f_\textrm{obj}$ is small $\iff$ the detected particles have a low transverse velocity.
		\item $f_\textrm{obj}$ should be continuous, even when all particles miss the detection laser.
		\item $f_\textrm{obj}$ should not get skewed by a few `bad' particles that are not detected.
	\end{enumerate}
	
	\noindent Requirements (1) and (2) are competing in the Liouville limit, since the particles that travel straight already occupy the phase space region centered around $(y,v_y)=(0,0)$. Consequently, it is hard to find an unambiguous objective function. In general different choices for the objective function are possible, each of them leading to a slightly different solution.
	
	Requirements (2-4) are naturally satisfied when we take as objective function the mean absolute transverse velocity $\langle |{v_y}| \rangle$ of the particles from the curved arm that can be detected by the laser. If we assume that the detection laser has a Gaussian profile with width $\sigma_\textrm{laser}$, we can write 
	
	\begin{equation}
		\label{eq:exp_vy}
		f_\textrm{obj}=\langle |{v_y}| \rangle = \frac{\sum_{i=1}^n w_i |v_{y,i}|}{\sum_{i=1}^n w_i} \quad , \text{with } \quad w_i = \exp(-\frac{1}{2}\left(\frac{y_i}{\sigma_\textrm{laser}}\right)^2),
	\end{equation}
	
	\noindent where $n$ is the number of particles in the simulated phase space, $i$ enumerates those particles, and $w_i$ is a weight proportional to the detection probability of each particle.
	
	Unfortunately, however, this function does not always satisfy requirement (1). A phase space distribution with small $|v_{y,i}|$ but large $y_i$ will still give a low value for $\langle |v_y| \rangle$. In this case, the merged beam runs parallel to the beam from the Stark decelerator but is offset in space, potentially limiting the overlap. We therefore choose to modify Eq.~\ref{eq:exp_vy} heuristically by dividing by the average weight one extra time, ensuring that low values for $y_i$ are favored. The objective function then becomes:
	
	\begin{equation}
		\label{eq:merged_guide_objective_function}
		f_\textrm{obj}(\vec{a}) = \frac{n\sum_{i=1}^n w_i |v_{y,i}|}{(\sum_{i=1}^n w_i)^2}.
	\end{equation}
	
	\noindent This objective function satisfies requirements (1-4), and will be used throughout. The optimization itself is carried out by MATLAB's built-in Nelder-Mead simplex algorithm called by the \verb|fminsearch| function~\cite{Lagarias1998}. Since many parameters are optimized simultaneously, this algorithm often converges to a local minimum. To explore a larger part of the parameter space the simplex could be reinitialized several times.

	\subsection{Numerically found solutions}
	
	\begin{figure}
		\includegraphics{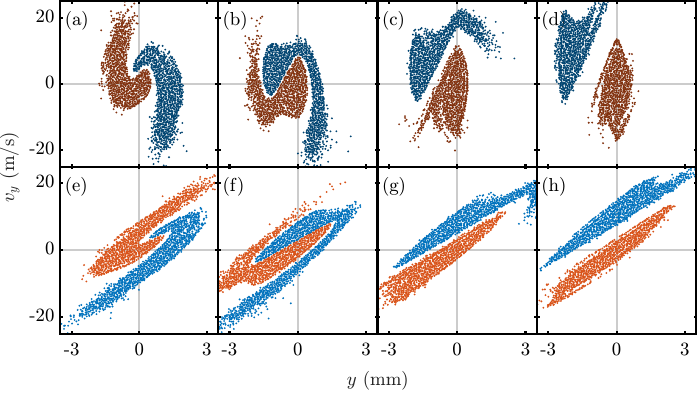}
		\vspace{-2mm} 
		\caption{
			\label{fig:merge_tstop_phasespaces}
			Phase space distributions $(y,v_y)$ for both packets produced by the finned merged guide parameterized by $\vec{a}_1$ (a,e), $\vec{a}_2$ (b,f), $\vec{a}_3$ (c,g) and $\vec{a}_4$ (d,h). Each design is terminated at a different $t_\text{stop}$, decreasing from left to right. See Tab.~\ref{table:merged_guide_solutions} for the exact parameters. Particles guided through the curved arm are displayed as blue dots, while the particles guided through the straight arm are displayed as orange dots. (a-d) phase space distribution at the exit of the merged guide. (e-h) phase space distribution at the plane of detection, \SI{65}{mm} behind the exit of the merged guide.
		}
	\end{figure}
	
	\begin{table}
		\caption{
			\label{table:merged_guide_solutions}
			Summary of the parameters used to define several finned merged guide designs. $i=0$ represents the initial guess, while $i=1-4$ represent designs found through numerical optimization, as described above. For each design, the score according to the objective function (Eq.~\ref{eq:merged_guide_objective_function}) is shown as well as the total length $L$ of the device.
		}
		\begin{ruledtabular}
			\begin{tabular}{c*{6}{r}*{4}{c}} 
				$i$ & \multicolumn{6}{c}{$\vec{a}_i$} & $\alpha_i$ ($^o$) & $t_{\text{stop},i}$ & $f_{\textrm{obj},i}$ & $L_i$ (mm) \\
				\hline
				0 & (0.84,&  4.47,& 23.79,&  57.28,& -37.83,& 38.41) & 7.9 & 1.00 &  3.7$\times10^6$ &  78\\ 
				1 & (0.60,&  4.34,& 33.95,&  89.75,& -23.13,& 53.95) & 6.0 & 1.00 & 12.3             & 151\\ 
				2 & (3.14,& 11.28,& 39.69,& 248.90,& -12.87,& 48.69) & 5.9 & 0.70 &  8.7             & 145\\
				3 & (3.08,& 11.17,& 40.39,& 245.15,& -13.26,& 52.36) & 5.6 & 0.55 & 15.0             &  98\\
				4 & (3.14,& 11.01,& 39.89,& 250.76,& -13.03,& 52.42) & 5.7 & 0.45 & 17.9             &  76\\
			\end{tabular}
		\end{ruledtabular}
	\end{table}
	
	We numerically optimized $\vec{a}$ starting with $\vec{a}_0$ and imposing a fixed starting angle of $\alpha=6^o$. The optimizer converged to a solution represented by vector $\vec{a}_1$, thereby decreasing the value of the objective function by over five orders of magnitude. The exact parameters can be found in Tab.~\ref{table:merged_guide_solutions}. Again, we evaluate the performance of this solution by looking at the $(y,v_y)$ phase space distribution of the resulting packets that are shown in Fig.~\ref{fig:merge_tstop_phasespaces}. It is seen that with solution $\vec{a}_1$ the two packets are dramatically closer at the exit of the guide, and the velocity distribution $v_y$ of the packet from the curved arm is nicely centered around $v_y=0$.  Upon exiting the guide, the two packets propagate together in free flight towards the plane of detection, yielding the phase space distributions shown in Fig.~\ref{fig:merge_tstop_phasespaces}(e). During this flight, both packets overlap spatially and are distributed near-symmetrically around the origin, although a strict boundary remains between them as imposed by Liouville's theorem. Instead of overlapping in both the spatial and velocity coordinates, the two packets swirl around one another.
	
	Solution $\vec{a}_1$ represents the optimal merging strategy within the geometrical constraints of the finned merged guide. On closer inspection, however, solution $\vec{a}_1$ may still be further improved. The phase space distribution of the two packets are still separated by a little distance throughout, whereas the Liouville theorem in principle allows for a situation in which both packets touch each other. In addition, it is seen that the secondary beam (blue dots) has already crossed the straight beam axis ($y=0$) and is now being focused back by the electrodes along the outer bend, i.e. it has already made a half-oscillation in phase space. The primary beam has formed a tail that has started to move upwards to relatively large values of $v_y$. 
	
	We empirically found that we can further improve on this design by shortening the length of the guide, limiting the amount of time both packets experience the potential well of the final hexapole. In our optimization routine, this can be achieved by omitting the final slices of the parametrization by limiting the range of $t$ to $t\in(-1,t_\text{stop})$. For each choice of $t_\text{stop}$ the numerical optimization was performed again, yielding new solutions. Three are presented here, with increasingly smaller $t_\text{stop}$: $\vec{a}_2$, $\vec{a}_3$ and $\vec{a}_4$ (see Tab.~\ref{table:merged_guide_solutions} for more details). The phase spaces at the exit of the merged guide are shown in Fig.~\ref{fig:merge_tstop_phasespaces}(b-d), while those at the plane of detection are shown in Fig.~\ref{fig:merge_tstop_phasespaces}(f-h).
	
	The design parametrized by $\vec{a}_2$ results in a phase space distribution that outperforms solution $\vec{a}_1$, as is also represented by the lower value for $f_\text{obj}$. The two packets are almost touching each other, and spatially overlap near perfectly during the free flight upon exiting the guide, with a minimum value for $v_y$ (and hence the collision energy). The total length of the design is \SI{145}{mm}, only \SI{6}{mm} shorter compared to the design pertaining to solution $\vec{a}_1$. 
	
	Yet, when considering the construction of a finned merged guide, additional requirements beyond the objective function may play a role. In our envisioned experiment, limiting the amount of collisions that can already take place inside the guide is essential. As mentioned before, these in-guide collisions can substantially complicate the interpretation of the scattering results, as the cross sections for the scattering between two dipolar molecules can increase by orders of magnitude in the presence of an external electric field. Ideally, the two packets should barely overlap upon exiting the guide, but have a phase space distribution such that full spatial overlap is only achieved in the free flight section downstream from the guide.  
	
	Referring back to Fig.~\ref{fig:merge_tstop_phasespaces}, solution $\vec{a}_2$ clearly offers the best overlap of both packets in phase space, yet, it scores less good with respect to our additional requirement of limiting the overlap inside the guide. A significant part of the secondary beam still crosses $y=0$ before exiting the guide, and the probability for collisions to occur inside the guide is relatively high. In this respect, solutions with an even lower value for $t_\text{stop}$ yield phase space distributions with a larger separation between the packets, together with an increased transverse velocity of the secondary beam (see solutions $\vec{a}_3$ and $\vec{a}_4$ in Fig.~\ref{fig:merge_tstop_phasespaces}). Although the values for $f_\text{obj}$ are less optimal for these solutions (but still reasonable), these solutions efficiently prevent in-guide collisions. They produce a phase space distribution for the secondary beam that lags behind the distribution of the primary packet ($y<0$), but with a positive mean transverse velocity $v_y>0$. During the free flight section between the guide and the detection point, the secondary beam overtakes the primary beam and full spatial overlap is achieved at the expense of a slightly larger mean collision energy.  In addition, these solutions substantially reduce the overall length of the guide, reducing the longitudinal spreading of the packets as they propagate through the setup.     
	
	All designs presented in this section are tailored to the experimental constraints incorporated into the optimization, including a \SI{370}{m/s} forward velocity. Considering these designs, and the rationale to gauge their performances, we chose to build the design represented by $\vec{a}_3$, as for our intended experiments this design offers a good compromise between efficient merging close to the Liouville limit, and more pragmatic problems related to the interpretation of the experimental scattering results we may obtain. The manufacturing of the finned merged guide according to this design, and its implementation into our existing crossed beam setup, will be described in detail in the next section.  
	
	\section{Implementation}\label{sec:implementation}
	
	\subsection{Manufacturing}
	\label{subsec:manufacturing}
	The numerically optimized design of the merged guide has a separation between both entrance quadrupoles of only \SI{6}{mm}, which is too small to interface the guide with a Stark decelerator or conventional beam sources. The guide was therefore extended at either entrance with ideal quadrupoles. On the straight arm, the electrodes were extended with a straight quadrupole of length \SI{107}{mm}, whereas the curved arm was extended with a curved quadrupole with \SI{641}{mm} radius of curvature, ending at \SI{15}{\degree} with respect to the straight axis. The latter value followed from extensive numerical optimization as described below in Sec.~\ref{subsec:interface}. These quadrupole extensions are fully integrated into the design of each electrode; further extensions are possible by placing additional beam manipulation elements before the merged guide entrances depending on requirements imposed by existing setups (see also Sec.~\ref{subsec:interface}).      
	
\begin{figure}
		\includegraphics[width=12cm]{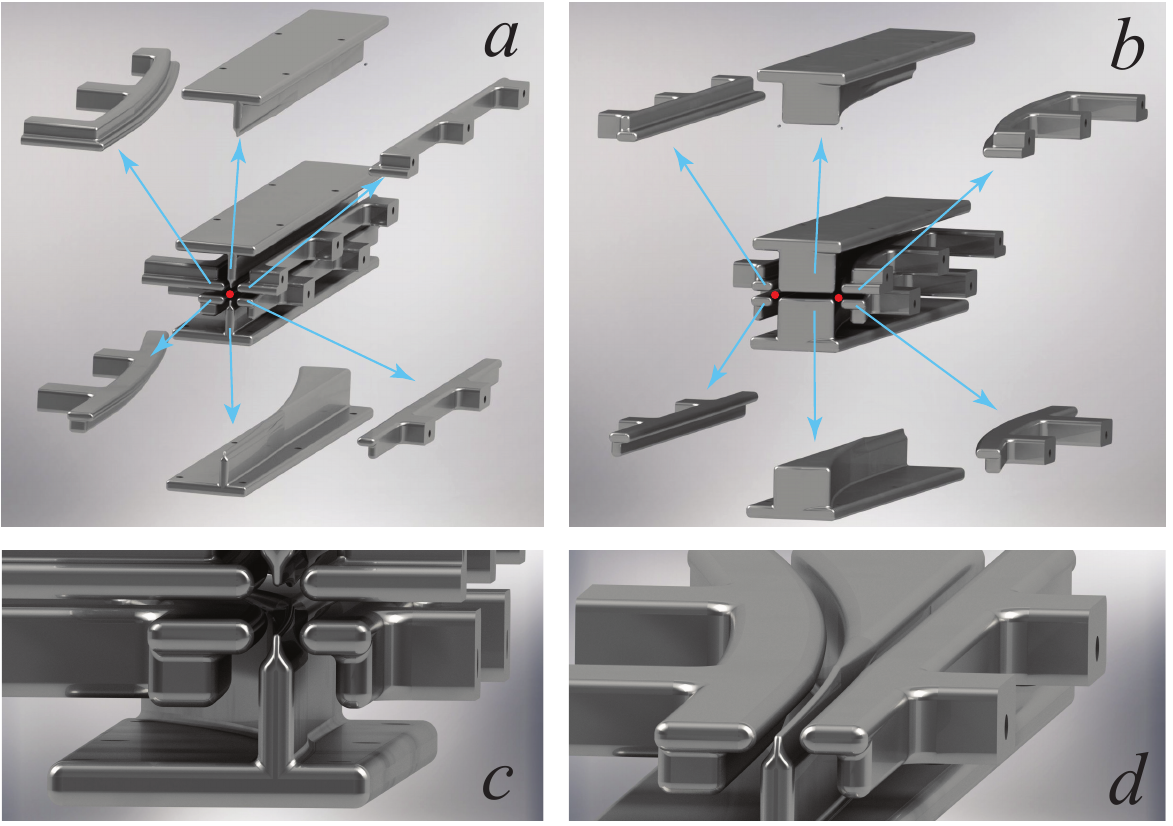}
		\caption{
			\label{fig:mergedblockexplosion}
			Exploded views of the six individual electrodes of the finned merged guide, showing the exit hexapole (a) and entrance quadrupoles (b) sides. The beam axis of the two entrances and exit are indicated by red dots. (c) Zoom-in of the exit hexapole side with a closer view on the finned electrodes. (d) Cut through the center plane showing the curvature of both arms.
		}
	\end{figure}

	Fig.~\ref{fig:mergedblockexplosion} shows an exploded view of the resulting six individual electrodes of the finned merged guide annex entrance quadrupoles. Panel (a) shows the exit side where the six electrodes of the hexapole are visible, whereas panel (b) shows the two quadrupole geometries at the entrance side. The beam axes at the entrance and exit are indicated by the red dots. For clarity, a zoom-in of the exit as well as a cut through the center plane of the construction are shown in panels (c) and (d), respectively. 
	
	The shape of the individual electrodes were first determined using the numerical optimization procedures described above. The 3D shape of all electrodes could be drawn in Autodesk Inventor 3D CAD software, using Eq.~\ref{eq:x_polynomial} and the geometrical constraints discussed in Sec.~\ref{subsec:finned_merged_guide}. The resulting geometry files are then directly read by a 5 axis CNC milling machine at Radboud University workshops. The electrodes were milled from aluminum 7075-T6, and manually polished after milling for excellent high voltage compatibility.

	\subsection{Mechanical support and alignment}
	In designing the mechanical support for the six electrodes, it is critical to 1) ensure proper alignment of the electrodes with respect to each other, and 2) to ensure excellent high voltage compatibility. For the former we impose an alignment accuracy of \SI{\sim 100}{\um}, whereas for the latter we know from experience that the most critical parts causing high voltage breakdown are insulators that separate mechanical parts placed at different potential from each other. 
	
\begin{figure}[b!]
		\includegraphics[width=12cm]{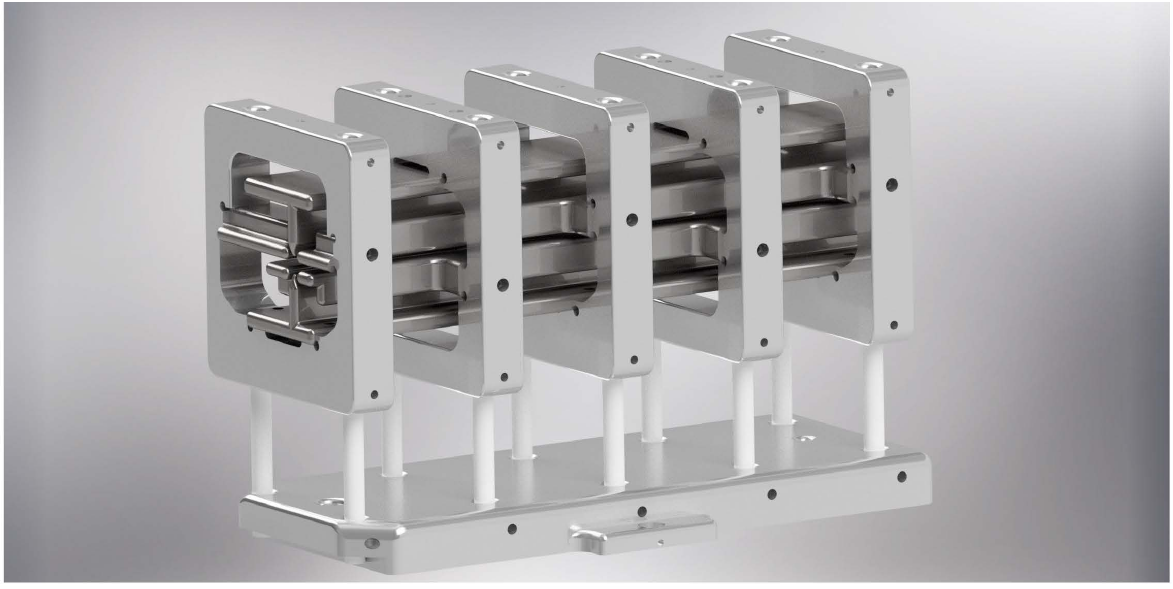}
		\caption{
			\label{fig:mergedblockassembly}
			Rendering of the merged guide assembly, showing the electrodes and their mounting brackets, ceramic support rods and base plate.
		}
	\end{figure}

	Both requirements are met using a mechanical electrode support concept as shown in Fig.~\ref{fig:mergedblockassembly}. The electrodes are clamped in squared aluminum brackets with rounded edges on all sides, preventing the need for insulators across the small distance between electrodes. Like for a standard quadrupole or hexapole, our finned merged guide is operated with only two independent voltages. One set of three electrodes share three mounting brackets, whereas the other set is mounted using only two brackets. The brackets have carefully designed and milled mounting planes, in which the electrodes lock into position, ensuring positioning of the electrodes with an accuracy defined by the high accuracy with which the mounting brackets are machined. The electrodes are tightened to the brackets using a screw from the side surfaces  of the brackets. Carefully placed venting holes ensure optimal vacuum compatibility.
	
	The two sets of three electrodes need to be aligned with respect to each other. This is ensured by locking the brackets onto a aluminum base plate using two ceramic (aluminum oxide; Al$_2$O$_3$) rods per bracket. Again, we solely rely on machine accuracy for all parts, with no freedom to manually align the structure. The holes supporting the Al$_2$O$_3$ rods are recessed with a relatively large radius of curvature to shield the triple points (interface between ceramic, metal and vacuum) that are known to easily initiate high voltage breakdown. A photograph of the assembly is shown in Fig.~\ref{fig:collage_real_merged_guide}(a).  
	
	The mounting concept as shown in Fig.~\ref{fig:mergedblockassembly} offers the additional advantage that the structure can be constructed and pre-tested with respect to mechanical and high voltage stability in a separate vacuum chamber, before installing it in the experiment. During high voltage conditioning tests, we have successfully applied a voltage of \SI{\pm13}{kV} between the two sets of electrodes for several hours, with currents below our detection limit of \SI{100}{nA}. The device has been routinely used in our experiments using a voltage of \SI{\pm12}{kV}, where additional high voltage switches were used to gate the high voltage pulses applied to the electrodes.

	\subsection{Interfacing the merged guide with a Stark decelerator beamline and conventional beam source}\label{subsec:interface}
	In our anticipated ND$_3$-ND$_3$ merged beam experiment, the straight and curved arms of the merged guide are fed by packets of ND$_3$ molecules produced by a Stark decelerator and conventional beam source, respectively. To accommodate sufficient space for the Stark decelerator, the separation between the two entrance quadrupoles needs to be further increased. This is accomplished by installing a straight quadrupole between the Stark decelerator and straight arm of the merged guide, and a curved hexapole in front of the curved arm. Unlike the integrated extensions mentioned in Sec.~\ref{subsec:manufacturing}, these extensions are separate electrostatic guides. 
	
\begin{figure}
		\includegraphics[width=12cm]{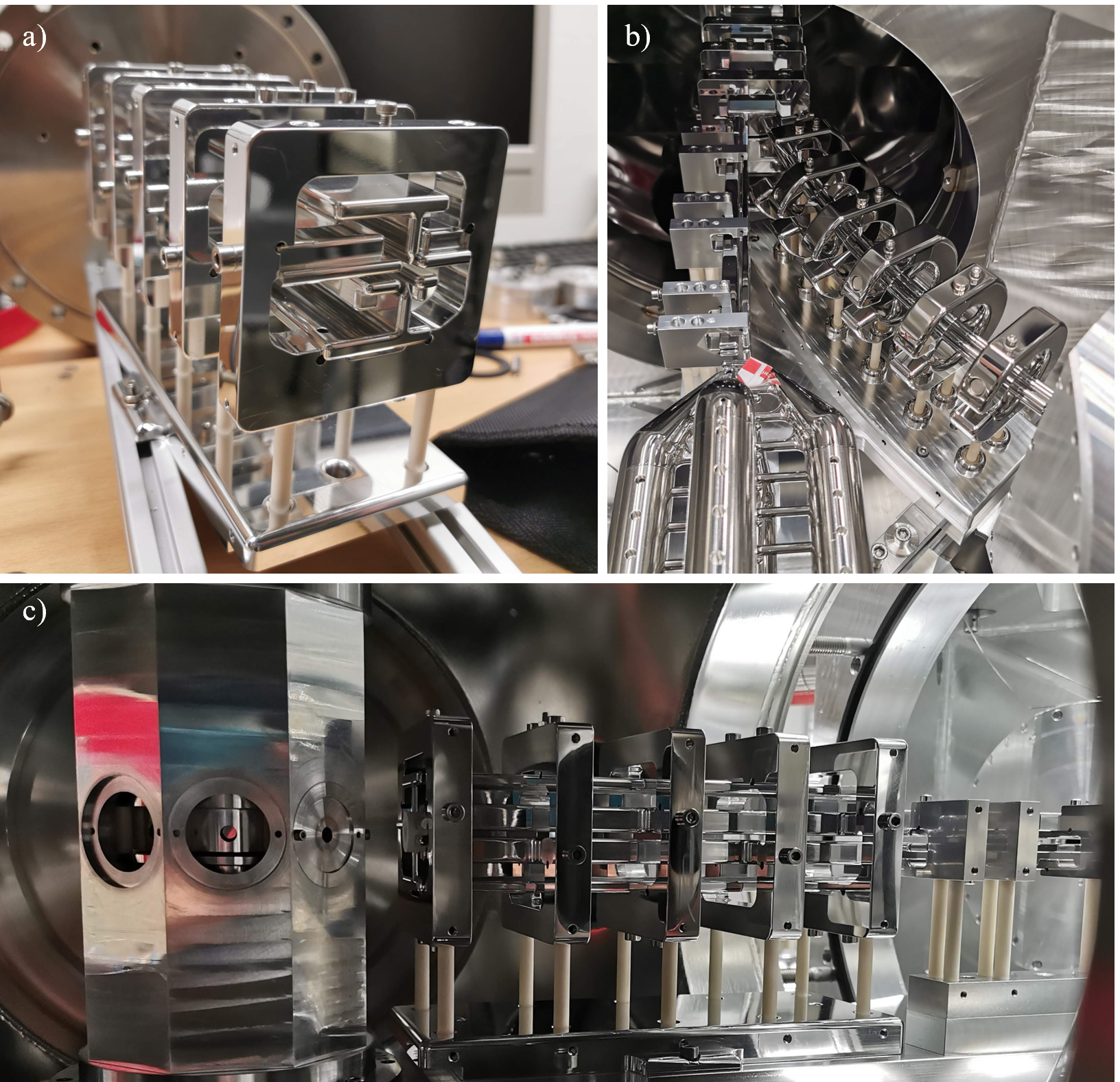}
		\caption{
			\label{fig:collage_real_merged_guide}
			Photos of the finished merged guide. (a) Photo from the exit hexapole, showing all six aluminum electrodes surrounding the beam axis. (b) Photo of the entire assembly, with a Stark decelerator feeding into a straight quadrupole followed by the straight arm of the merged guide. From the right, a curved hexapole guides the molecules to the entrance of the curved arm of the merged guide. (c) Photo of the last part of the assembly. From right to left, the straight quadrupole, merged guide and VMI lens.
		}
	\end{figure}

	The straight quadrupole has a length of \SI{20}{cm}, in which the quadrupole field is generated by the rounded edges of four plate electrodes. Like in the merged guide design, these electrodes are supported by aluminum brackets mounted on a base plate using ceramic feet. A photo of part of this quadrupole is shown in Fig.~\ref{fig:collage_real_merged_guide}(b) and (c). Only modest electric fields are required to transport the ND$_3$ molecules emerging from the Stark decelerator to the merged guide, and the quadrupole was operated by applying voltages of only \SI{\pm1.5}{kV} to the electrodes.    
	
	On the curved side, one could consider installing either a curved quadrupole or a hexapole. A quadrupole offers a linear dependence of the electric field as a function of the displacement from the beam axis, whereas for a hexapole this dependence is quadratic. Consequently, molecules with a predominantly linear Stark effect (such as ND$_3$) experience a linear restoring force resulting in a near-perfect oscillatory motion in a hexapole, whereas molecules with a predominantly quadratic Stark effect (such as NH$_3$) are best manipulated using a quadrupole. As our main objective involves ND$_3$-ND$_3$ collisions (with a possible future extension of a second Stark decelerator to load the curved arm), we opted for installing a curved hexapole. Using NH$_3$ in the curved arm is then still possible, but its trajectories are less nicely described. 
	
	As a quadrupole and hexapole have quite different phase space acceptances, care must be taken to design the curved hexapole such that the packet of ND$_3$ molecules smoothly transitions from the curved hexapole into the curved quadrupole entrance of the merged guide. From extensive numerical optimizations, we found that a curved hexapole with \SI{641}{mm} radius of curvature, spanning a \SI{25}{\degree} segment of a circle, is optimal. These parameters result in the tight focusing of a ND$_3$ packet traveling with our target velocity of \SI{370}{m/s} into the quadrupole entrance by the hexapole, i.e., the phase space distribution has a narrow spatial but large transverse velocity spread upon entrance of the merged guide. As the quadrupole has a large velocity acceptance, this distribution is nicely matched to the acceptance region of the quadrupole, resulting in the least amount of heating. The ND$_3$ molecules traveled through the curved hexapole at a distance of \SI{1.54}{mm} from the geometric center due to the centrifugal force. We aligned this equilibrium orbit with the geometric center of the quadrupole entrance. A photograph of the curved hexapole is seen in Fig.~\ref{fig:collage_real_merged_guide}(b).
	
	\begin{figure}
		\includegraphics{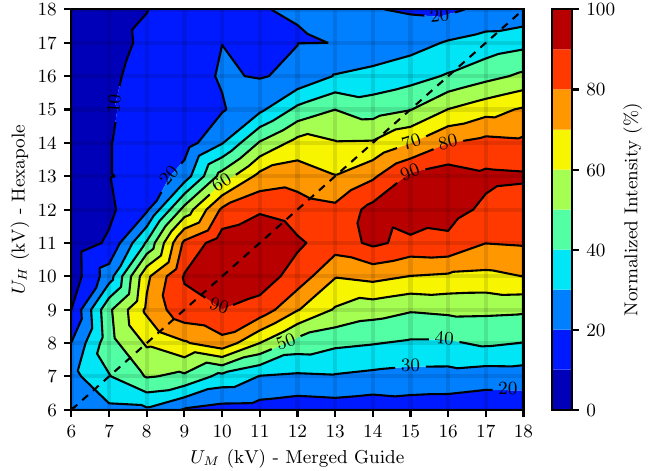}
		\vspace{-1mm}
		\caption{
			\label{fig:2D_voltage_hex-merge}
			Normalized simulated peak intensity of the secondary beam as a function of the voltages applied to the merged guide $U_M$ and curved hexapole $U_H$. The dashed line shows the diagonal, $U_M=U_H$.
		}
	\end{figure} 
	
	Numerical simulations were performed to find the optimal voltages to operate the curved hexapole. Fig.~\ref{fig:2D_voltage_hex-merge} shows the normalized expected peak intensity as a function of the voltages applied to the merged guide and curved hexapole. It is seen that the maximum intensity is found near the diagonal, i.e., optimal performance is expected when the curved hexapole and merged guide share the same voltages. Deviations from this diagonal are only found for voltages exceeding \SI{\pm11}{kV}. We therefore electrically connected the electrodes of the curved hexapole to the electrodes of the merged guide, such that both devices can share a single set of high voltage power supplies and switches.

	\section{Results}\label{sec:results}
	\subsection{Trajectory simulations}	
	In order to verify the performance of the merged guide in conjunction with the rest of the setup, we performed full 3D numerical trajectory simulations. Realistic time-dependent phase space distributions of ND$_3$ molecules were generated for each source, mimicking the molecular beam pulses on both sides of the experiment. To assess the performance in the most critical situation, both beams travel at matching velocities around \SI{370}{m/s}. These distributions were then propagated through each consecutive manipulation element along each beamline, i.e., a Stark decelerator, quadrupole, and merged guide for the straight arm, and a curved hexapole in conjunction with the merged guide for the curved arm. Collisions between the two packets were simulated using the same method as Tang \emph{et al.} ~\cite{Tang2023}, taking the phase space overlap of both packets emerging from the merged guide into account. For this, we only considered the \SI{65}{mm} free flight section from the guide exit to the laser ionisation region, and ignored the possibility that both packets may already partially overlap before exiting the merged guide.   
	
	The results are summarized in Tab.~\ref{table:merged_guide_comparison}. In this table, we list the relevant performance for different hypothetical versions of the experimental setup, such as the relative densities of the straight ($I_1$) and curved ($I_2$) beams, the collision energy ($E$), collision energy resolution ($\sigma_E$) and the relative number of collisions ($I_{col}$) that may be expected from the simulated overlap of both packets. For ease of comparison between the different configurations, we normalize the intensity of the Stark decelerated beam, after propagation in free flight to the detection region, to 100. For the curved beamline, the intensity that is found if the curved hexapole as used by Tang \emph{et al.} is used throughout, is also normalized to 100. We note that the suggestion that both these reference beams have equal intensity is not true; from experiments it follows that the packet of molecules exiting the curved beamline has approximately a factor \num{\sim10} to \num{100} higher density than the Stark-decelerated packet. 
	
	\begin{table}
		\caption{
			\label{table:merged_guide_comparison}
			Simulated benchmark parameters for several possible merged beam setups using two beams 1 and 2. The proposed setup refers to decelerator, quadrupole guide, merged guide for beamline 1 and curved hexapole, merged guide for beamline 2. Both beams contain ND$_3$ molecules, except in the last row, where NH$_3$ is used in the simulation of beamline 2. The two beams travel at matching velocities around \SI{370}{m/s}. $I_i$ shows the relative peak beam intensity, assuming a laser volume of \SI{4}{mm} wide, \SI{1}{mm} high. The average collision energy $\langle E \rangle$, collision energy resolution $\sigma_E$ and detected collision intensity $I_\text{col}$ were found by simulating collisions between the phase spaces as in Ref.~\onlinecite{Tang2023}, assuming a uniform ICS. $\sigma_E$ is calculated as the square root of the variance of the collision energy distribution.
		}
		\begin{ruledtabular}
			\begin{tabular}{*{7}{c}}
				Beamline 1 & Beamline 2  & $I_{1}$ & $I_{2}$  & $\langle E\rangle$ & $ \sigma_E$ & $I_\mathrm{col}$ \\ 
				& & (\si{\percent}) & (\si{\percent}) & (cm$^{-1}$) & (cm$^{-1}$) & (rel.) \\
				\hline   
				Decelerator         & none 		& 100 & - & - & - & -  \\
				Decelerator         & Curved hexapole  & 15 & 100 & 0.25 & 0.14 & 1.0 \\
				Proposed & Proposed  & 555  & 32 & 0.17 & 0.10 & 23 \\
				Proposed & Proposed, NH$_3$ & 555  & 56  & 0.15 & 0.08 & 29 \\
			\end{tabular}
		\end{ruledtabular}
	\end{table}
	
	The second line of Tab.~\ref{table:merged_guide_comparison} pertains to the situation where no merged guide is used, i.e., this is the experimental configuration as used by Tang \emph{et al.} for the NO-ND$_3$ system. It is observed that if this configuration is used for ND$_3$-ND$_3$, about \SI{85}{\percent} of the ammonia molecules that exit the Stark decelerator are deflected by the curved hexapole. The particles that do pass the curved hexapole are transversely heated significantly, occupying a ring in phase space that surrounds the secondary beam. This results in a rather poor beam overlap and a relatively high minimal collision energy, that is about four times larger than the minimal collision energy obtained by Tang \emph{et al.} for NO-ND$_3$. For the curved side, the harmonic potential of the curved hexapole is ideal for molecules with a linear Stark effect like ND$_3$, such that the beam can be focused towards the interaction region, resulting in an intense secondary beam.
	
	The third line in Tab.~\ref{table:merged_guide_comparison} corresponds to the situation where the finned merged guide, including the quadrupole and curved hexapole to load the guide from the straight and curved arms, respectively, is installed. A large increase in density is observed in the straight arm. With the quadrupole and merged guide in place, particles are efficiently transversely focused upon exiting of the Stark decelerator, yielding a large intensity increase for the primary beam. The intensity of the curved secondary beam however, is reduced by \SI{70}{\percent}. From the simulations, we conclude that this loss is mainly caused by that fact that molecules are first guided by a hexapole field and then loaded into a quadrupole field in the first segment of the merged guide. Consequently, molecules do not experience a harmonic potential throughout, resulting in less than ideal focusing properties and reduction in density. Yet, the overall performance of this arrangement for ND$_3$-ND$_3$ is dramatically superior to a strategy of performing a ND$_3$-ND$_3$ collision experiment using the original arrangement that was designed and ideal for NO-ND$_3$. The minimum collision energy of \SI{0.17}{\per\cm} is only about a factor 2 larger than the energy that could be achieved for NO-ND$_3$, as the merged guide inevitably compromises the phase space overlap of both beams that can fundamentally be achieved. By simply comparing the product of the two beam intensities between the second and third line, one would expect a factor of $(5.55\times0.32)/(0.15\times1.00)\approx12$ increase in collision signal. However, this value is even exceeded, as a factor of 23 is predicted by the simulations. This underlines the superior overlap of the two beams achieved by the merged guide, allowing for a larger probability for collisions to occur during their flight from the guide exit to the detection region.
	
	For the last line of the table, the curved arm was loaded with NH$_3$ molecules rather than  ND$_3$, which will be the preferred configuration to probe ND$_3$-NH$_3$ collisions in our experiments.  It is observed that the curved arm of the merged guide performs a factor 1.7 better with NH$_3$ compared to ND$_3$. This is perhaps surprising, as the merged guide was optimized for ND$_3$ in our simulations. A full analysis of this observation is beyond the scope of this work, but we hypothesize that the quadratic Stark shift of NH$_3$ that is present up to rather large electric fields reduces the loss of molecules during the transition from a hexapole to a quadrupole configuration. The resulting minimal achievable collision energy amounts to \SI{0.15}{\per\cm}, whereas the gain in expected collision signal is slightly below a factor 30.
	
	Based on the predicted performances of the merged guide and Stark decelerator assembly as summarized in Tab.~\ref{table:merged_guide_comparison}, the expected beam densities and collision signal levels are sufficient to perform ND$_3$-ND$_3$ and ND$_3$-NH$_3$ scattering experiments at collision energies down to \SI{0.15}{\per\cm}. Preliminary collision experiments for these systems using the merged guide are presented in Sec.~\ref{sec:collisions}.

	\subsection{Experimental verification}
	The performance of the merged guide was experimentally tested by probing the peak intensity of the ND$_3$ molecules emerging from either the straight or curved arm using standard 2+1 REMPI at a distance of \SI{65}{mm} from the exit of the guide. The straight arm was loaded with a packet of ND$_3$ molecules from a \SI{2.6}{m} long Stark decelerator as described in detail elsewhere~\cite{Tang2023}. The curved arm was loaded with a conventional molecular beam. Both molecular beam pulses were generated using a room-temperature NPV~\cite{Yan:RSI84:023102} by entraining approximately \SI{2}{\percent} ND$_3$ in a carrier gas at a backing pressure of \SI{1}{\bar}, and skimmed using a \SI{3}{mm} diameter skimmer placed approximately \SI{15}{cm} from the nozzle orifice. 
	
	For the straight arm, the peak signal intensity was measured as a function of ND$_3$ velocity and voltages applied to the decelerator, quadrupole and merged guide, both for the situation where the quadrupole and merged guide were switched on or off. Overall, good qualitative agreement with the simulations were obtained (data not shown), although the signal gain when applying voltages to the quadrupole and merged guide were over-estimated in the simulations by about a factor 1.5. This signal gain is very sensitive to the exact laser detection volume, and we therefore consider this quantitative disagreement not unreasonable.      
	
	The curved arm was tested using ND$_3$ seeded in Xe throughout, as the design of the merged guide was optimized for a fixed velocity of \SI{370}{m/s}. The peak signal intensity was measured as a function of the voltage applied to the curved hexapole and merged guide. It was observed that maximum transmission is obtained for voltages of \SI{\pm11}{kV}, which was in quantitative agreement with the simulations. We limited our experimental tests to below \SI{\pm12}{kV} to avoid a possible electrical breakdown.
	
	The performance of the curved arm was further investigated by measuring full arrival time distributions of the ND$_3$ molecules arriving in the detection region. This distribution will generally not directly represent the velocity distribution of the original molecular beam, as the transmittance of the curved hexapole and merged guide is expected to critically depend on the velocity of the molecules. Molecules that are too fast will not follow the curvature, whereas molecules that are too slow will have sub-optimal focusing properties within the structure. In particular, efficient transfer of molecules from the curved hexapole to the merged quadrupole guide is only possible if the (transverse) phase space distribution of the molecules evolves as assumed in our simulations. 
	
	Consequently, only a narrow velocity class will be transmitted. We can take advantage of these effects, and develop experimental probes to critically assess the performance of the curved arm by switching off the voltages on the guide prematurely. The idea is that with an early interruption of the guiding fields, only a subset of the molecules will arrive at the detector after propagating in free flight. These molecules originate from a narrow region in phase space (negative position but positive velocity coordinate), such that measuring the arrival time distribution for different switch off times directly probes the phase space distribution of the molecules within the guide. We introduce the parameter $t_{\text{off}}$ to define the switching off with respect to the timing of the detection laser, which is kept fixed for all measurements and defines $t=0$. The arrival time distribution is then effectively measured by scanning the timing at which the valve fires. 
	
	\begin{figure}
		\includegraphics{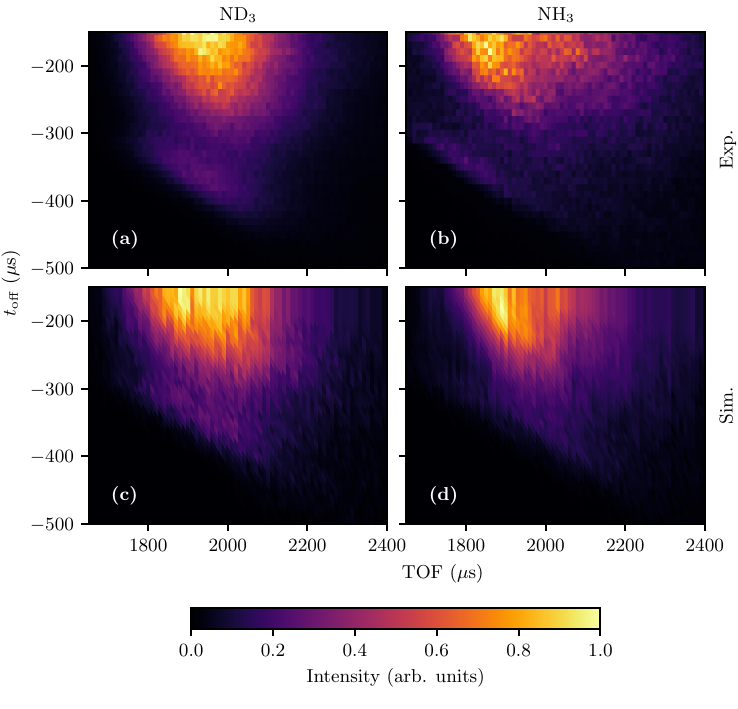}
		\caption{
			\label{fig:merge_guide_2Dtofs}
			Experimental (top) and simulated (bottom) time of flight (TOF) scans of the curved beampath for ND$_3$ (left) or NH$_3$ (right) while switching the merging electrodes off prematurely at $t_\text{off}$. $t_\text{off}$ is relative to TOF, i.e. the moment of detection.
		}
	\end{figure}
	
	\begin{figure}
		\includegraphics{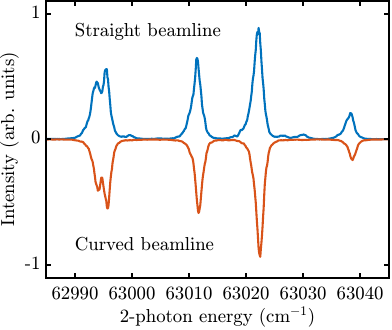}
		\caption{
			\label{fig:ND3_REMPI_spectra}
			Rotational 2+1 REMPI spectra of ND$_3$ molecules emerging from both beamlines (see text). All five main transitions originate from the $1_1^-$ state. The spectrum corresponding to the straight beamline (top) additionally shows three weak features that originate from the $2_2^-$ state.
		}
	\end{figure} 
	
	Fig.~\ref{fig:merge_guide_2Dtofs}(a) shows the arrival time distributions as a function of $t_{\text{off}}$ as a 2D plot, together with the expected distributions from simulations in panel (c). It is seen that qualitative agreement is obtained between the experimental and simulated distributions. Maximum intensity is observed for $t_{\text{off}} \geq \SI{-200}{\us}$, as the molecules then experience fields throughout their entire journey through the guide. When the guide is switched off earlier, the arrival time distribution narrows, in agreement with our interpretation that only a subset of the molecular distribution can reach the detector in free flight. An interesting revival is observed for $t_{\text{off}}\sim \SI{-400}{\us}$, which is also observed in the simulations. The reduced intensity observed at $t_{\text{off}}\sim \SI{-350}{\us}$ likely results from a part of the phase space where the molecules exit the merged guide more diverging.  
	
	Similar experiments and simulations were conducted for NH$_3$ molecules that pass through the curved arm, see Fig.~\ref{fig:merge_guide_2Dtofs}(b,d). As explained earlier, NH$_3$ will behave differently from ND$_3$ in the hexapole and merged guide, as the large inversion splitting causes a second order Stark effect up to relatively high fields. Nevertheless, beams of NH$_3$ can still be used in the curved arm, and we in fact observed a slightly higher signal intensity for NH$_3$ compared to ND$_3$, in agreement with Tab.~\ref{table:merged_guide_comparison}. The measured arrival time distributions for NH$_3$ are generally narrower compared to ND$_3$, in qualitative agreement with simulations.    
	
	To probe the quantum state purity of the ND$_3$ molecules emerging from the merged guide, we recorded 2+1 REMPI spectra in the $B(\nu'_2=5) \leftarrow X(\nu_2=0)$ region~\cite{Ashfold:JCP89:1754}, where $\nu_2$ indicates the vibrational quantum number for the umbrella mode of ND$_3$. Fig.~\ref{fig:ND3_REMPI_spectra} shows the REMPI spectra recorded for both the straight (upper trace) and the curved (lower trace) beamlines. To record these spectra, the same conditions for the laser were used as during scattering measurements, i.e., we optimize for high detection efficiency by chosing laser powers to just saturate the transitions. Consequently, the spectral resolution in the measured spectra is limited by power broadening. From the spectra, it can be seen that the molecules originating from both beamlines are rotationally cold and near-exclusively populate the low-field-seeking $J_K^p=1_1^-$ state. In the spectrum pertaining to the straight beamline, three weak spectral features can be discerned. These features can be attributed to the  $2_2^-$ first rotationally excited state, which lies \SI{14.5}{\per\cm} above the $1_1^-$ ground state and has a small initial population in the beam. This state is low-field seeking, and molecules in this level are focused by the beam manipulators. Population in this state is not observed for the curved beamline, most likely because the Stark effect of the $2_2^-$ state is not ideal to properly guide the molecules through the curved beam elements.

	\section{Preliminary collision experiments}\label{sec:collisions}
	To demonstrate the feasibility of performing low-energy collision experiments with our new merged guide, we performed preliminary collision experiments using the ND$_3$-ND$_3$ system. This system is of particular interest, as recently an interesting scattering mechanism was predicted to occur in collisions between two polar molecules, where each molecule possesses near-degenerate rotational levels with opposite parity~\cite{Tang2023}. 
	
	We tested the ability to observe scattering signal for ND$_3$-ND$_3$ collisions under the most difficult conditions, i.e., when both beams travel with matching velocities resulting in the lowest collision energy. The premise is that if we can unequivocally observe scattering signal under these conditions, scanning towards higher energies should then be feasible as well, as the kinematic conditions and beam densities will become more favorable as the collision energy is increased. 
	
	We operated the Stark decelerator to produce packets of ND$_3$ molecules with velocities ranging between \num{375} and \SI{400}{m/s}. The curved arm was loaded with ND$_3$ at a fixed velocity of \SI{375}{m/s}. Isotope labeling was used in order to spectroscopically separate signals from the Stark decelerator ($^{15}$ND$_3$) and curved beam ($^{14}$ND$_3$). Inelastic collisions are probed by recording the collision induced population increase in the $1_1^+$ level of the packet of 
	$^{15}$ND$_3$ molecules emerging from the decelerator using 2+1 REMPI. The integral ion signal on a selected part of the MCP was recorded, without attempting to record images using velocity map imaging.   
	
	Signals were recorded by scanning the velocities of the Stark decelerated packet using a repeated cycle addressing six different velocities. During this cycle, for each velocity the signal is recorded for three different configurations of the experiment, averaging over 40 consecutive shots of the experiment per configuration. In the first, both beams are in full overlap, such that collisions can occur. In the second, only the straight beam is operated, generating a background signal in the $1_1^+$ detection channel originating from this beam alone. In the third, the $1_1^+$ background signal is recorded by operating exlusively the curved beam. This cycle is then repeated 265 times, after which the number of recorded ions per velocity and per configuration is binned.    
	
	\begin{figure}
		\includegraphics{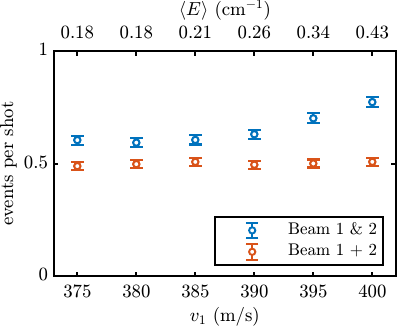}
		\caption{
			\label{fig:ND3-ND3_collisions}
			Average number of detected events per shot for $^{15}$ND$_3$-$^{14}$ND$_3$ collisions using the merged guide. Both partners are prepared in the $1_1^-$ state, while the $1_1^+$ state of $^{15}$ND$_3$ is probed for several primary beam velocities, $v_1$. The velocity of the secondary beam $v_2$ is fixed at 375 m/s. The mean collision energy is indicated at the top. The detection rate with both beams on simultaneously (blue data points) exceeds the sum of either beam separately (red data points). Vertical error bars represent the statistical errors at \SI{95}{\percent} confidence interval. The difference between the curves results from population transfer induced by inelastic collisions.
		}
	\end{figure} 
	
	Fig.~\ref{fig:ND3-ND3_collisions} presents the resulting average recorded events per shot of the experiment. The combined background of both individual beams is plotted (red data points), together with the signals recorded when both beams are brought in overlap (blue data points). Clearly, the blue data points are well above the red ones throught the energy range probed. The collision signal levels are small ($\sim 0.1$ detected collision event per shot), but are discerned with statistical relevance within a reasonable signal accumulation time. We conclude that with the merged guide presented here, low-energy collision experiments using two dipolar molecules are feasible at collision energies down to 0.17~cm$^{-1}$. 
	
	Is is noted that the combined background signal is dominated by the Stark decelerated beam, which accounts for about 90 $\%$ of the recorded background signal. This background signal mainly originates from $^{15}$ND$_3$ molecules that reside in the $1_1^+, M=0$ level in the original beam. Molecules in this state experience, in first order, zero Stark shift and propagate through the apparatus without being affected by electric fields. The curved arm hardly contributed to the background level, as i) only trace amounts of $^{15}$ND$_3$ are present in this beam, and ii) molecules in the $1_1^+, M=0$ level will not follow the curvature of the hexapole and guide. The background in the straight beam could potentially be reduced substantially by designing a merged guide that curves both beams (as in the original design by Osterwalder \emph{et al.}), but this would compromise our desire to propagate the Stark decelerated beam on a straight axis throughout the entire beamline. 
	
	\section{Conclusions}
	We have developed an electrostatic guide that allows for the optimal merging of two beams of polar molecules. Based on a previous design by Osterwalder \emph{et al.}, we optimized the shape of the electrodes using extensive numerical simulations to merge the two beams with optimal overlap in phase space allowed by the Liouville theorem, i.e., yielding optimal probability for scattering at the lowest possible collision energy. Furthermore, the design allows for merging of both beams in a field-free region upon exiting the guide, facilitating the interpretation of experimental scattering results. The merged guide was successfully used to observe scattering for the $^{15}$ND$_3$-$^{14}$ND$_3$ system at collision energies down to \SI{0.18}{\per\cm}. The merged guide presented here offers a powerful method to experimentally probe low-energy collisions between two dipolar molecules with unprecedented accuracy and in hitherto unexplored energy regimes, and can help us to unravel the peculiar and highly interesting low-energy collision properties of dipolar molecules.   
	
	\begin{acknowledgments}
		This work is part of the research program of the Netherlands Organization for Scientific
		Research (NWO). S.Y.T.v.d.M. acknowledges support from the European Research
		Council (ERC) under the European Union’s Horizon 2020 Research and Innovation
		Program (Grant Agreement No. 817947 FICOMOL). We thank A. Osterwalder (EPFL Lausanne) for carefully reading the manuscript, and for stimulating discussions. We thank N. Janssen and T. Kuijs for excellent technical support.
	\end{acknowledgments}
	
	\newpage
	\bibliographystyle{plainnat}
	\bibliography{bibliography.bib}

\providecommand{\noopsort}[1]{}\providecommand{\singleletter}[1]{#1}%
\begin{thebibliography}{40}
\providecommand{\natexlab}[1]{#1}
\providecommand{\url}[1]{\texttt{#1}}
\expandafter\ifx\csname urlstyle\endcsname\relax
  \providecommand{\doi}[1]{doi: #1}\else
  \providecommand{\doi}{doi: \begingroup \urlstyle{rm}\Url}\fi

\bibitem[Amarasinghe and Suits(2017)]{Amarasinghe2017}
Chandika Amarasinghe and Arthur~G Suits.
\newblock Intrabeam scattering for ultracold collisions.
\newblock \emph{J. Phys. Chem. Lett.}, 8\penalty0 (20):\penalty0 5153--5159,
  2017.
\newblock \doi{10.1021/acs.jpclett.7b02378}.

\bibitem[Amarasinghe et~al.(2020)Amarasinghe, Li, Perera, Besemer, Zuo, Xie,
  van~der Avoird, Groenenboom, Guo, K{\l}os, and Suits]{Amarasinghe2020}
Chandika Amarasinghe, Hongwei Li, Chatura~A. Perera, Matthieu Besemer, Junxiang
  Zuo, Changjian Xie, Ad~van~der Avoird, Gerrit~C. Groenenboom, Hua Guo, Jacek
  K{\l}os, and Arthur~G. Suits.
\newblock State-to-state scattering of highly vibrationally excited {NO} at
  broadly tunable energies.
\newblock \emph{Nat. Chem.}, 12\penalty0 (6):\penalty0 528--534, may 2020.
\newblock \doi{10.1038/s41557-020-0466-8}.

\bibitem[Ashfold et~al.(1988)Ashfold, Dixon, Little, Stickland, and
  Western]{Ashfold:JCP89:1754}
M.~N.~R. Ashfold, R.~N. Dixon, N.~Little, R.~J. Stickland, and C.~M. Western.
\newblock The $\widetilde{B} {}^{1}{E}''$ state of ammonia: sub-{D}oppler
  spectroscopy at vacuum ultraviolet energies.
\newblock \emph{J. Chem. Phys.}, 89:\penalty0 1754--1761, 1988.
\newblock \doi{10.1063/1.455715}.

\bibitem[Chefdeville et~al.(2013)Chefdeville, Kalugina, van~de Meerakker,
  Naulin, Lique, and M.]{Chefdeville:Science341:06092013}
S.~Chefdeville, Y.~Kalugina, S.~Y.~T. van~de Meerakker, C.~Naulin, F.~Lique,
  and Costes M.
\newblock Observation of partial wave resonances in low-energy {O$_2$-H$_2$}
  inelastic collisions.
\newblock \emph{Science}, 341\penalty0 (6150):\penalty0 1094--1096, 2013.
\newblock \doi{10.1126/science.1241395}.

\bibitem[Dahl(2000)]{Dahl2000}
David~A Dahl.
\newblock {SIMION} for the personal computer in reflection.
\newblock \emph{Int. J. Mass Spectrom.}, 200\penalty0 (1–3):\penalty0 3--25,
  December 2000.
\newblock ISSN 1387-3806.
\newblock \doi{10.1016/S1387-3806(00)00305-5}.

\bibitem[Deng and Yin(2007)]{Deng2007}
Lianzhong Deng and Jianping Yin.
\newblock Beam splitter for guided polar molecules with a {Y}-shaped charged
  wire.
\newblock \emph{Opt. Lett.}, 32\penalty0 (12):\penalty0 1695, jun 2007.
\newblock \doi{10.1364/ol.32.001695}.

\bibitem[Deng et~al.(2011)Deng, Liang, Gu, Hou, Li, Xia, and Yin]{Deng2011}
Lianzhong Deng, Yan Liang, Zhenxing Gu, Shunyong Hou, Shengqiang Li, Yong Xia,
  and Jianping Yin.
\newblock Experimental demonstration of a controllable electrostatic molecular
  beam splitter.
\newblock \emph{Phys. Rev. Lett.}, 106\penalty0 (14):\penalty0 140401, apr
  2011.
\newblock \doi{10.1103/PhysRevLett.106.140401}.

\bibitem[Dulitz et~al.(2020)Dulitz, van~den Beld-Serrano, and
  Stienkemeier]{Dulitz2020}
Katrin Dulitz, Marco van~den Beld-Serrano, and Frank Stienkemeier.
\newblock Single-source, collinear merged-beam experiment for the study of
  reactive neutral--neutral collisions.
\newblock \emph{J. Phys. Chem. A}, 124\penalty0 (17):\penalty0 3484--3493,
  2020.
\newblock \doi{10.1021/acs.jpca.0c00608}.

\bibitem[Engelhart et~al.(2015)Engelhart, Wagner, Johnsen, Wodtke, and
  Schäfer]{Engelhart2015}
Daniel~P. Engelhart, Roman J.~V. Wagner, Peter~C. Johnsen, Alec~M. Wodtke, and
  Tim Schäfer.
\newblock Adsorbate enhancement of electron emission during the quenching of
  metastable {CO} at metal surfaces.
\newblock \emph{Phys. Chem. Chem. Phys.}, 17\penalty0 (17):\penalty0
  11540--11545, 2015.
\newblock \doi{10.1039/C5CP01255D}.

\bibitem[Gawlas and Hogan(2019)]{Gawlas2019}
K~Gawlas and SD~Hogan.
\newblock Rydberg-state-resolved resonant energy transfer in cold
  electric-field-controlled intrabeam collisions of {NH$_3$} with {R}ydberg
  {He} atoms.
\newblock \emph{J. Phys. Chem. Lett.}, 11\penalty0 (1):\penalty0 83--87, 2019.
\newblock \doi{10.1021/acs.jpclett.9b03290}.

\bibitem[Gordon and Osterwalder(2017)]{Gordon2017:3D}
Sean~DS Gordon and Andreas Osterwalder.
\newblock {3D}-printed beam splitter for polar neutral molecules.
\newblock \emph{Phys. Rev. Appl.}, 7\penalty0 (4):\penalty0 044022, 2017.
\newblock \doi{10.1103/PhysRevApplied.7.044022}.

\bibitem[Gordon et~al.(2018)Gordon, Omiste, Zou, Tanteri, Brumer, and
  Osterwalder]{Gordon2018}
Sean~DS Gordon, Juan~J Omiste, Junwen Zou, Silvia Tanteri, Paul Brumer, and
  Andreas Osterwalder.
\newblock Quantum-state-controlled channel branching in cold {Ne (${}^3$P$_2$)
  + Ar} chemi-ionization.
\newblock \emph{Nat. Chem.}, 10\penalty0 (12):\penalty0 1190--1195, 2018.
\newblock \doi{10.1038/s41557-018-0152-2}.

\bibitem[Henson et~al.(2012)Henson, Gersten, Shagam, Narevicius, and
  Narevicius]{Henson:Science338:234}
A.~B. Henson, S.~Gersten, Y.~Shagam, J.~Narevicius, and E.~Narevicius.
\newblock Observation of resonances in penning ionization reactions at
  sub-kelvin temperatures in merged beams.
\newblock \emph{Science}, 338\penalty0 (6104):\penalty0 234--238, 2012.
\newblock \doi{10.1126/science.1229141}.

\bibitem[Herbers et~al.(2022)Herbers, Caris, Kuijpers, Grabow, and van~de
  Meerakker]{Herbers2022}
S.~Herbers, Y.~M. Caris, S.~E.~J. Kuijpers, J.-U. Grabow, and S.~Y.~T. van~de
  Meerakker.
\newblock Efficient transfer of inversion doublet populations in deuterated
  ammonia using adiabatic rapid passage.
\newblock \emph{Mol. Phys.}, page e2129105, 2022.
\newblock \doi{10.1080/00268976.2022.2129105}.

\bibitem[Herbers et~al.(2024)Herbers, Sustar, Kuijpers, Sparling, and van~de
  Meerakker]{Herbers2024}
S.~Herbers, P.~Sustar, S.~Kuijpers, C.~Sparling, and S.~Y.~T. van~de Meerakker.
\newblock Efficient transfer of population between rotational levels in
  deuterated ammonia using {THz} transitions.
\newblock \emph{Mol. Phys.}, 2024.
\newblock \doi{10.1080/00268976.2024.2353333}.
\newblock in press.

\bibitem[Hogan(2016)]{Hogan2016}
Stephen~D Hogan.
\newblock Rydberg-{S}tark deceleration of atoms and molecules.
\newblock \emph{EPJ Tech. Instrum.}, 3:\penalty0 1--50, 2016.
\newblock \doi{10.1140/epjti/s40485-015-0028-4}.

\bibitem[Ketterle and Pritchard(1992)]{Ketterle:PRA46:4051}
W.~Ketterle and D.~E. Pritchard.
\newblock Atom cooling by time-dependent potentials.
\newblock \emph{Phys. Rev. A}, 46:\penalty0 4051--4054, 1992.
\newblock \doi{10.1103/PhysRevA.46.4051}.

\bibitem[Klein et~al.(2017)Klein, Shagam, Skomorowski, {\.Z}uchowski, Pawlak,
  Janssen, Moiseyev, van~de Meerakker, van~der Avoird, Koch, and
  Narevicius]{Klein:NatPhys13:35}
Ayelet Klein, Yuval Shagam, Wojciech Skomorowski, Piotr~S {\.Z}uchowski,
  Mariusz Pawlak, Liesbeth~MC Janssen, Nimrod Moiseyev, Sebastiaan~YT van~de
  Meerakker, Ad~van~der Avoird, Christiane~P Koch, and Edvardas Narevicius.
\newblock Directly probing anisotropy in atom--molecule collisions through
  quantum scattering resonances.
\newblock \emph{Nat. Phys.}, 13\penalty0 (1):\penalty0 35, 2017.
\newblock \doi{10.1038/nphys3904}.

\bibitem[Koller et~al.(2022)Koller, Jung, Phrompao, Zeppenfeld, Rabey, and
  Rempe]{Koller2022}
Manuel Koller, Florian Jung, Jindaratsamee Phrompao, Martin Zeppenfeld,
  IM~Rabey, and Gerhard Rempe.
\newblock Electric-field-controlled cold dipolar collisions between trapped
  {CH$_3$F} molecules.
\newblock \emph{Phys. Rev. Lett.}, 128\penalty0 (20):\penalty0 203401, 2022.
\newblock \doi{10.1103/PhysRevLett.128.203401}.

\bibitem[Lagarias et~al.(1998)Lagarias, Reeds, Wright, and
  Wright]{Lagarias1998}
Jeffrey~C. Lagarias, James~A. Reeds, Margaret~H. Wright, and Paul~E. Wright.
\newblock Convergence properties of the {N}elder--{M}ead simplex method in low
  dimensions.
\newblock \emph{SIAM J. Optim.}, 9\penalty0 (1):\penalty0 112--147, jan 1998.
\newblock \doi{10.1137/S1052623496303470}.

\bibitem[Lavert-Ofir et~al.(2014)Lavert-Ofir, Shagam, Henson, Gersten, K{\l}os,
  {\.Z}uchowski, Narevicius, and Narevicius]{Lavert-Ofir:NatChem6:332}
E.~Lavert-Ofir, Y.~Shagam, A.~B. Henson, S.~Gersten, J.~K{\l}os, P.~S.
  {\.Z}uchowski, J.~Narevicius, and E.~Narevicius.
\newblock Observation of the isotope effect in sub-kelvin reactions.
\newblock \emph{Nat. Chem.}, 6:\penalty0 332--335, 2014.
\newblock \doi{10.1038/nchem.1857}.

\bibitem[Liouville(1838)]{Liouville:JMPA3:342}
Joseph Liouville.
\newblock Sur la th\'{e}orie de la variation des constantes arbitraires.
\newblock \emph{J. Math. Pures Appl.}, 3:\penalty0 342, 1838.
\newblock URL \url{http://www.numdam.org/item/JMPA_1838_1_3__342_0.pdf}.

\bibitem[Margulis et~al.(2022)Margulis, Paliwal, Skomorowski, Pawlak,
  {\.Z}uchowski, and Narevicius]{Margulis2022}
Baruch Margulis, Prerna Paliwal, Wojciech Skomorowski, Mariusz Pawlak, Piotr~S
  {\.Z}uchowski, and Edvardas Narevicius.
\newblock Observation of the p-wave shape resonance in atom-molecule
  collisions.
\newblock \emph{Phys. Rev. Research}, 4\penalty0 (4):\penalty0 043042, 2022.
\newblock \doi{10.1103/PhysRevResearch.4.043042}.

\bibitem[Note1()]{Note1}
Note1.
\newblock It is noted that the electrode shape is symmetric, but the applied
  voltages are anti-symmetric.

\bibitem[Note2()]{Note2}
Note2.
\newblock The distance between the electrode surfaces will be slightly smaller
  when measured along the surface normal instead of this cross section.

\bibitem[Osterwalder(2015)]{Osterwalder:EPJ-TI2:10}
Andreas Osterwalder.
\newblock Merged neutral beams.
\newblock \emph{EPJ Tech. Instrum.}, 2:\penalty0 10, 2015.
\newblock \doi{10.1140/epjti/s40485-015-0022-x}.

\bibitem[O’Connor et~al.(2015)O’Connor, Urbain, St{\"u}tzel, Miller,
  de~Ruette, Garrido, and Savin]{OConnor2015}
A.~P. O’Connor, Xavier Urbain, J.~St{\"u}tzel, K.~A. Miller, N.~de~Ruette,
  M.~Garrido, and Daniel~Wolf Savin.
\newblock Reaction studies of neutral atomic {C} with {H$_3^+$} using a
  merged-beams apparatus.
\newblock \emph{Astrophys. J. Suppl. Ser.}, 219\penalty0 (1):\penalty0 6, 2015.
\newblock \doi{10.1088/0067-0049/219/1/6}.

\bibitem[Palmer and Hogan(2017)]{Palmer2017}
J.~Palmer and S.~D. Hogan.
\newblock Experimental demonstration of a {R}ydberg-atom beam splitter.
\newblock \emph{Phys. Rev. A}, 95\penalty0 (5):\penalty0 053413, may 2017.
\newblock \doi{10.1103/PhysRevA.95.053413}.

\bibitem[Perreault et~al.(2017)Perreault, Mukherjee, and
  Zare]{Perreault:science358:356}
W.~E. Perreault, N.~Mukherjee, and R.~N. Zare.
\newblock Quantum control of molecular collisions at 1 kelvin.
\newblock \emph{Science}, 358:\penalty0 356, 2017.
\newblock \doi{10.1126/science.aao3116}.

\bibitem[Phaneuf et~al.(1999)Phaneuf, Havener, Dunn, and
  M{\"u}ller]{Phaneuf1999}
R.~A. Phaneuf, C.~C. Havener, G.~H. Dunn, and A.~M{\"u}ller.
\newblock Merged-beams experiments in atomic and molecular physics.
\newblock \emph{Rep. Prog. Phys.}, 62\penalty0 (7):\penalty0 1143, 1999.
\newblock \doi{10.1088/0034-4885/62/7/202}.

\bibitem[Plomp et~al.(2020)Plomp, Gao, and van~de Meerakker]{Plomp2020}
Vikram Plomp, Zhi Gao, and Sebastiaan Y.~T. van~de Meerakker.
\newblock A velocity map imaging apparatus optimised for high-resolution
  crossed molecular beam experiments.
\newblock \emph{Mol. Phys.}, page e1814437, sep 2020.
\newblock \doi{10.1080/00268976.2020.1814437}.

\bibitem[Riedel et~al.(2011)Riedel, Hoekstra, J\"ager, Gilijamse, van~de
  Meerakker, and Meijer]{Riedel:EPJD65:161}
J.~Riedel, S.~Hoekstra, W.~J\"ager, J.~J. Gilijamse, S.~Y.~T. van~de Meerakker,
  and G.~Meijer.
\newblock Accumulation of {S}tark-decelerated {NH} molecules in a magnetic
  trap.
\newblock \emph{Eur. Phys. J. D}, 65:\penalty0 161, 2011.
\newblock \doi{10.1140/epjd/e2011-20082-7}.

\bibitem[Sawyer et~al.(2007)Sawyer, Lev, Hudson, Stuhl, Lara, Bohn, and
  Ye]{Sawyer:PRL98:253002}
B.~C. Sawyer, B.~L. Lev, E.~R. Hudson, B.~K. Stuhl, M.~Lara, J.~L. Bohn, and
  J.~Ye.
\newblock Magnetoelectrostatic trapping of ground state {OH} molecules.
\newblock \emph{Phys. Rev. Lett.}, 98:\penalty0 253002, 2007.
\newblock \doi{10.1103/PhysRevLett.98.253002}.

\bibitem[Shimizu et~al.(2003)Shimizu, Che, Hashinokuchi, Fukuyama, Suzui,
  Watanabe, and Kasai]{Shimizu2003}
Y.~Shimizu, D.-C. Che, M.~Hashinokuchi, T.~Fukuyama, M.~Suzui, M.~Watanabe, and
  T.~Kasai.
\newblock Electrostatic hexapole state selector: Honeycomb field for
  integrating intensity of oriented molecular beams.
\newblock \emph{Rev. Sci. Instrum.}, 74\penalty0 (8):\penalty0 3749--3752, July
  2003.
\newblock ISSN 1089-7623.
\newblock \doi{10.1063/1.1593793}.

\bibitem[Smith(2006)]{Smith2006}
Ian W.~M. Smith.
\newblock Reactions at very low temperatures: Gas kinetics at a new frontier.
\newblock \emph{Angewandte Chemie International Edition}, 45\penalty0
  (18):\penalty0 2842--2861, 2006.
\newblock \doi{10.1002/anie.200502747}.

\bibitem[Tang et~al.(2023)Tang, Besemer, Kuijpers, Groenenboom, van~der Avoird,
  Karman, and van~de Meerakker]{Tang2023}
Guoqiang Tang, Matthieu Besemer, Stach Kuijpers, Gerrit~C. Groenenboom,
  Ad~van~der Avoird, Tijs Karman, and Sebastiaan Y.~T. van~de Meerakker.
\newblock Quantum state-resolved molecular dipolar collisions over four decades
  of energy.
\newblock \emph{Science}, 379\penalty0 (6636):\penalty0 1031--1036, 2023.
\newblock \doi{10.1126/science.adf9836}.

\bibitem[Tanteri(2021)]{Tanteri:thesis:2021}
Silvia Tanteri.
\newblock \emph{Studies of Cold Chemistry in Crossed and Merged Neutral Beams.}
\newblock {Ph.D.} thesis, Ecole Polytechnique F{\'e}d{\'e}rale de Lausanne,
  2021.
\newblock URL
  \url{https://epfl.swisscovery.slsp.ch/permalink/41SLSP_EPF/1g1fbol/alma99116780513005516}.

\bibitem[van~de Meerakker et~al.(2012)van~de Meerakker, Bethlem, Vanhaecke, and
  Meijer]{Meerakker:CR112:4828}
Sebastiaan Y.~T. van~de Meerakker, Hendrick~L. Bethlem, Nicolas Vanhaecke, and
  Gerard Meijer.
\newblock Manipulation and control of molecular beams.
\newblock \emph{Chem. Rev.}, 112:\penalty0 4828--4878, 2012.
\newblock \doi{10.1021/cr200349r}.

\bibitem[Yan et~al.(2013)Yan, Claus, van Oorschot, Gerritsen, Eppink, van~de
  Meerakker, and Parker]{Yan:RSI84:023102}
B.~Yan, P.~F.~H. Claus, B.~G.~M. van Oorschot, L.~Gerritsen, A.~T. J.~B.
  Eppink, S.~Y.~T. van~de Meerakker, and D.~H. Parker.
\newblock A new high intensity and short-pulse molecular beam valve.
\newblock \emph{Rev. Sci. Instrum.}, 84:\penalty0 023102, 2013.
\newblock \doi{10.1063/1.4790176}.

\bibitem[Zhelyazkova et~al.(2020)Zhelyazkova, Martins, Agner, Schmutz, and
  Merkt]{Zhelyazkova2020}
Valentina Zhelyazkova, Fernanda~BV Martins, Josef~A Agner, Hansj{\"u}rg
  Schmutz, and Fr{\'e}d{\'e}ric Merkt.
\newblock Ion-molecule reactions below 1 {K}: strong enhancement of the
  reaction rate of the ion-dipole reaction {He$^+$ + CH$_3$F}.
\newblock \emph{Phys. Rev. Lett.}, 125\penalty0 (26):\penalty0 263401, 2020.
\newblock \doi{10.1103/PhysRevLett.125.263401}.

\end{thebibliography}
	
\end{document}